\documentclass{aa} 

\usepackage{graphicx}
\usepackage{setspace}
\usepackage{natbib}
\usepackage{longtable}
\usepackage{amssymb}
\usepackage[varg]{txfonts}
\usepackage[usenames]{color}

\usepackage{hyperref}

\usepackage[T1]{fontenc}
\usepackage{url}
\bibpunct{(}{)}{;}{a}{}{,}

\newcommand{\eg}{e.g.}
\newcommand{\ie}{i.e.}

\newcommand{\teff}{$T_{{\rm eff}}$}

\newcommand{\vr}{$v_{\rm rad}$}
\newcommand{\logg}{{\rm log\,{\it g}}}

\newcommand{\ms}{m\,s$^{\rm -1}$}
\newcommand{\kms}{km\,s$^{\rm -1}$}
\newcommand{\Msun}{M$_{\sun}$}

\newcommand{\bdor}{$\beta$~Dor}
\newcommand{\lcar}{$\ell$~Car}
\newcommand{\dcep}{$\delta$~Cep}

\newcommand{\rvc}{$RV_{\rm cc-c}$}
\newcommand{\rvg}{$RV_{\rm cc-g}$}
\newcommand{\rvbg}{$RV_{\rm cc-2g}$}
\newcommand{\coo}{$\mathcal{C}^{\rm o}$}
\newcommand{\asym}{$A_{\rm cc}$}
\newcommand{\sncc}{SNR$_{\rm CCF}$}
\newcommand{\dcore}{$\Delta_{\rm core}$}
\newcommand{\dwing}{$\Delta_{\rm wings}$}

\newcommand{\all}{all}
\newcommand{\weak}{weak}
\newcommand{\medm}{medium}
\newcommand{\deep}{deep}

\newcommand{\gr}{green}
\newcommand{\red}{red}
\newcommand{\blue}{blue}

\begin{document} %

\title{Consistent radial velocities of classical Cepheids from the cross-correlation technique\thanks{Partly based on observations made with ESO telescopes at Paranal and La Silla observatories under program IDs: 072.D-0419, 073.D-0136, 091.D-0469(A), 097.D60150(A) and 190.D-0237 for HARPS data; 098.D-0379(A), 0100.D-0397(A) and 0101.D-0551(A) for UVES data; and 073.D-0072(A), 074.D-0008(B), 075.D-0676(A), 084.B-0029(A) and 087.D-0603(A) for FEROS data. Partly based on observations made with the SOPHIE spectrograph at the Observatoire de Haute-Provence (CNRS, France), under program IDs PNPS.FRAN (12B), PNPS.GALL (13A, 14A, 15A), PNPS.KERV (13B, 16B, 17B), and PNPS.BORG (18A). Partly based on observations made with the CORALIE spectrograph on the Euler telescope (Swiss Observatory) at La Silla, Chile, under program numbers 1 and 756. Partly based on observations collected at the Telescope Nazionale Galileo in the framework of the OPTICON proposal 2015B/15 for HARPS-North data.}}

\author{S.~Borgniet \inst{1} \and P.~Kervella \inst{1} \and N.~Nardetto \inst{2} \and A.~Gallenne \inst{2,3} \and A.~M\'erand \inst{4} \and R.I.~Anderson \inst{4} \and J.~Aufdenberg \inst{5} \and L.~Breuval \inst{1} \and W.~Gieren \inst{6,7} \and V.~Hocd\'e \inst{2} \and B.~Javanmardi \inst{1} \and E.~Lagadec \inst{2} \and G.~Pietrzy\'nski \inst{6,8} \and B.~Trahin \inst{1}}

\institute{
LESIA, Observatoire de Paris, Universit\'e PSL, CNRS, Sorbonne Universit\'e, Universit\'e de Paris, 5 Place Jules Janssen, 92195 Meudon, France
\and
Universit\'e C\^ote d'Azur, OCA, CNRS, Lagrange, Parc Valrose, B\^at. Fizeau, 06108 Nice Cedex 02, France
\and
European Southern Observatory, Alonso de C\'{o}rdova 3107, Casilla 19001, Santiago, Chile
\and
European Southern Observatory, Karl-Schwarzschild-Str. 2, D-85748 Garching, Germany
\and
Embry-Riddle Aeronautical University, Physical Sciences Department, 600~S Clyde Morris Boulevard, Daytona Beach, FL 32114, USA
\and 
Universidad de Concepci\'on, Departamento de Astronom\'ia, Casilla 160-C, Concepci\'on, Chile
\and
Millenium Institute of Astrophysics, Av. Vicuna Mackenna 4860, Santiago, Chile
\and
Nicolaus Copernicus Astronomical Centre, Polish Academy of Sciences, Bartycka 18, PL-00-716 Warszawa, Poland
}

\offprints{simon.borgniet@obspm.fr}
\date{Received
         ...; accepted ...}

\abstract
   {Accurate radial velocities (\vr) of Cepheids are mandatory within the context of Cepheid distance measurements via the Baade-Wesselink technique. The most common \vr~derivation method consists in cross-correlating the observed stellar spectra with a binary template and measuring a velocity on the resulting mean profile. Yet, for Cepheids and other pulsating stars, the spectral lines selected within the template as well as the way of fitting the cross-correlation function (CCF) have a direct and significant impact on the measured \vr.}
   {Our first aim is to detail the steps to compute consistent CCFs and \vr~of Cepheids. Next, this study aims at characterising the impact of Cepheid spectral properties and \vr~computation methods on the resulting line profiles and \vr~time series.}
   {We collected more than 3900 high-resolution spectra from seven different spectrographs of 64 Classical Milky Way (MW) Cepheids. These spectra were normalised and standardised using a single, custom-made process on pre-defined wavelength ranges. We built six tailored correlation templates selecting un-blended spectral lines of different depths based on a synthetic Cepheid spectrum, on three different wavelength ranges from 3900 to 8000 \AA. Each observed spectrum was cross-correlated with these templates to build the corresponding CCFs, adopted as the proxy for the spectrum mean line profile. We derived a set of line profile observables as well as three different \vr~measurements from each CCF and two custom proxies for the CCF quality and amount of signal.}
   {This study presents a large catalogue of consistent Cepheid CCFs and \vr~time series. It confirms that each step of the process has a significant impact on the deduced \vr: the wavelength, the template line depth and width, and the \vr~computation method. The way towards more robust Cepheid \vr~time series seems to go through steps that minimise the asymmetry of the line profile and its impact on the \vr. Centroid or first moment \vr, that exhibit slightly smaller amplitudes but significantly smaller scatter than Gaussian or biGaussian \vr, should thus be favoured. Stronger or deeper spectral lines also tend to be less asymmetric and lead to more robust \vr~than weaker lines.} 
   {}

\keywords{Techniques: spectroscopy -- techniques: radial velocities -- Stars: variable: Cepheids -- Stars: atmospheres}
   
\authorrunning{S. Borgniet et al.}
\titlerunning{Consistent radial velocities of classical Cepheids.}
   
\maketitle
\section{Introduction}

Cepheids are essential extra-galactic distance standards as their period of pulsation ($P$) correlates tightly with their absolute luminosity, a (mostly empirical) relationship known for more than a century as the Leavitt law, or the Cepheid Period-Luminosity (hereafter $P$-L) relationship \citep{leavitt12}. They have thus vastly contributed to precision cosmology, and especially to the measurement of the Hubble constant $H_{\rm 0}$ \citep{freedman10,riess11,riess16,riess18}. Their high-amplitude pulsations, that generally follow a relatively clear pattern, also make them targets of choice for spectroscopic studies looking for a better understanding of the stellar structure. 

Radial velocities (\vr) of Cepheids have thus long been of high interest. First, they allow to detect and characterise Cepheid stellar companions (\ie~single-line spectroscopic binaries or SB1s), that have revealed themselves to be widespread \citep[see \eg][]{evans15,anderson15,anderson16a,gallenne13,gallenne15,gallenne18b,gallenne18a}. The most common application of Cepheid \vr~is their use for the measurement of the Cepheid distances through the Parallax-of-Pulsation method, also known as the Baade-Wesselink (BW) technique \citep{lindemann18,baade26,wesselink46}. The BW method allows to derive directly the distance in a quasi-geometrical way through the ratio of the Cepheid linear radius variation ($\Delta R$, measured spectroscopically) over the angular diameter variation $\Delta \theta$ \citep[see recent examples in \eg][]{kervella04,gallenne12,gallenne17,merand15,nardetto18}. 

The linear radius variation $\Delta R$ is assumed to be proportional to the \vr~curve integrated over the pulsation phase. To understand how, it is first necessary to remind what exactly represents the radial velocity \citep[see also][on the definition]{lindegren03} in the context of a pulsating star and especially a Classical Cepheid. Cepheids are radial pulsators with a given photospheric pulsation velocity ($v_{\rm puls}$) at a given phase of the pulsation period. $v_{\rm puls}$ is a true physical quantity, whereas we measure the resulting Doppler shift of the received stellar light integrated over the stellar disk and projected on our line of sight. This Doppler shift is measured either on a single spectral line or on the cross-correlation function (CCF) of the full spectrum with a correlation template. Each region of the stellar disk is more or less Doppler-shifted and contributes more or less to the total spectrum depending on its position on the disk and on the limb-darkening. As a consequence, the Doppler shift of the spectrum integrated over the stellar disc (and the corresponding \vr) is mitigated by a certain amount compared to the physical $v_{\rm puls}$. This amount is called the projection factor (or $p$-factor) and it rounds up all the various sources of physical and spectral variability \citep{burki82,nardetto07}. It is also the key parameter of classical BW methods as it is fully degenerate with the distance \citep{merand15}.

Another consequence of the radial projection effect is that each line (and thus the global line profile) of the total Cepheid spectrum is intrinsically asymmetric \citep{sabbey95,nardetto06}. This has a strong and variable effect on the measurement of the Doppler shift of the line, depending on the measurement method. Furthermore, it is now common knowledge that Cepheid spectral lines are shifted differently depending on their formation region or depth, \ie~that an atmospheric velocity gradient is present. Thus, any Cepheid \vr~measurement is affected, depending on which spectral lines are used \citep{nardetto07,anderson16}. In short, to each \vr~measurement corresponds a different $p$-factor (and thus potentially a different distance estimation). Finally, at a deeper precision level, additional phenomena might introduce uncertainty on the \vr~and $p$-factors: shockwaves \citep{nardetto18}, convective blueshift \citep{nardetto08}, or cycle-to-cycle variability \citep{anderson16}.

Cepheid studies do not always clarify how the \vr~were computed. Furthermore and most often, the template used to cross-correlate the spectra is not described or provided, except in very few cases \citep{brahm17}. However, given the context developed above, authors should always make clear what spectral lines were used to derive the \vr~and how the \vr~were computed, as stressed by \cite{anderson17}. 

\begin{figure}[ht!]
\centering
\includegraphics[width=1.\hsize]{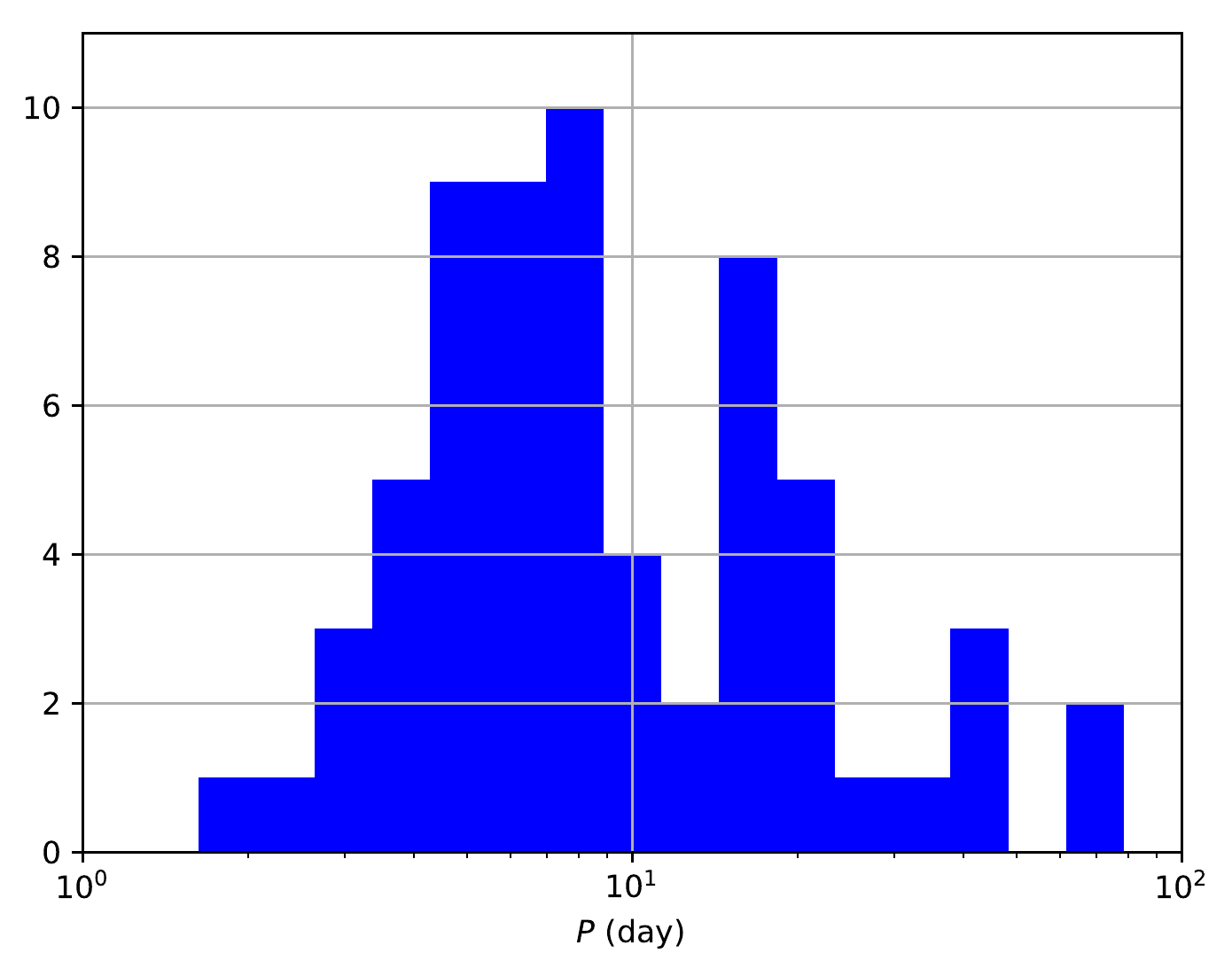}
\caption{Pulsation period ($P$) distribution of our Cepheid sample.}
\label{fig:P_dist}
\end{figure}

We present here a consistent spectroscopic survey of 64 Classical Milky Way (MW) Cepheids. It includes for each target the CCFs, various line profile observables, and several \vr~time series built from different correlation templates and computation methods. In Sect.~\ref{sect:survey}, we describe our Cepheid sample, the observations performed and the data used in this study. We introduce the principles of our framework and the main outputs in Sect.~\ref{sect:method}. We then apply our method to derive a consistent set of CCFs, observables and \vr~for our full Cepheid sample. We discuss and characterise the results of the survey in Sect.~\ref{sect:results}. We finally conclude on the perspectives and possible applications of this survey in Sect.~\ref{sect:conclu}.

\section{Survey description}\label{sect:survey}

\subsection{Sample}
Our sample is made up of 64 Classical Galactic Cepheids with pulsation periods in the range $\sim$2 to $\sim$68 days. We provide the full detail of our sample in Table~\ref{tab:sample} (Appendix~\ref{appdix:sample}). We selected our targets based on the number of available high-resolution spectra (and the number of corresponding distinct observation epochs) that we could recover either from new observations or from spectroscopic archives (see below). We selected only targets with a good enough sampling and coverage (\ie~$\gtrsim$ 20\%) of the pulsation phase. We show the pulsation period distribution of our sample in Fig.~\ref{fig:P_dist}. Most of our targets have a period between a few and 20 days, but we also include a small number of Cepheids with periods up to several tens of days in order to scan as well as possible the whole Cepheid period range.

\subsection{Data}

We gathered a total of 3919 high-resolution (40~000~$\lesssim$ $R = \lambda / \mathrm{d} \lambda$ $\lesssim$~115~000) spectra acquired with seven different echelle spectrographs. These instruments are located in both hemispheres and allow us to cover a wide wavelength range (from the near ultraviolet to the near infrared), depending on their respective characteristics (Table~\ref{tab:list_spectro}). 

\renewcommand{\arraystretch}{1.25}
\begin{table*}[t!]
\caption{Spectrographs implemented in this study. Column 3 gives the spectral resolution $R$ and col.~4 the instrumental (systematic) \vr~uncertainty. Col.~5 gives the wavelength range(s) $\Delta\lambda$ on which we standardised and reduced the spectra (see text). Columns~6 and 7 give the number of observed Cepheids $N_{\star}$ and the total number of acquired spectra $N_{\rm sp}$ for each instrument.}
\label{tab:list_spectro}
\begin{center}
\begin{tabular}{l l | c c | c c c}\\
\hline
\hline
Spectrograph  & Observatory  & Spectral res.                      & \vr                & $\Delta\lambda$   & $N_{\star}$ & $N_{\rm sp}$ \\
              &              & $R = \lambda / \mathrm{d} \lambda$ & precision$^{(a)}$   &                   &            &      \\
              &              &                                    &   (\ms)            &                   &            &      \\
\hline
HARPS        & ESO La Silla, 3.6m telescope            & $\sim$115~000$^{(b)}$ & 0.5             & \gr              &  36        & 429  \\
FEROS        & ESO La Silla, 2.2m telescope            & $\sim$48~000         & 21              & \gr, \red        &  25        & 93   \\
UVES         & ESO VLT, UT2$^{(c)}$ 8m telescope        & $\sim$70~000         & 25              & \blue, \red      &  24        & 2331 \\
CORALIE      & La Silla$^{(d)}$, Euler 1.2m telescope   & $\sim$55~000         & 3               & \gr              &  22        & 483 \\
SOPHIE       & OHP$^{(e)}$, 1.93m telescope             & $\sim$75~000$^{(b)}$  & 5               & \gr              &  30        & 370 \\
HARPS-North  & Roque$^{(f)}$, TNG$^{(g)}$ 3.6m telescope & $\sim$115~000        & 0.5             & \gr              &  1         & 103  \\
HERMES       & Roque$^{(f)}$, Mercator 1.2m telescope   & $\sim$85~000         & 2.5             & \gr, \red        &  3         & 110  \\
\hline
\hline
\end{tabular}
\end{center}
{\it (a)} According to the literature. {\it (b)} When used in the High-Resolution mode. {\it (c)} Unitary Telescope 2 (Kueyen) at ESO Very Large Telescope Observatory (Cerro Paranal, Chile). {\it (d)} Swiss Observatory at Cerro La Silla, Chile. {\it (e)} Observatoire de Haute-Provence, France. {\it (f)} Observatorio del Roque de los Muchachos, La Palma, Canary Islands. {\it (g)} Telescopio Nazionale Galileo. 
\end{table*}
\renewcommand{\arraystretch}{1}

\subsubsection{New observations}
We detail here our new observations that were not published before. Our observational strategy was adapted to maximise the pulsation phase coverage for each target.

\paragraph{SOPHIE --} From June 2013 to August 2018, we observed 30 northern Cepheids with the fibre-fed SOPHIE spectrograph \citep{bouchy06} mounted on the 1.93m telescope at Observatoire de Haute-Provence (France). We acquired 296 spectra in the High-Resolution mode ($R \sim$75~000). In addition, we gathered 74 High-Efficiency ($R \sim$40~000) spectra from the public SOPHIE archive\footnote{\url{http://atlas.obs-hp.fr/sophie/}.}. Exposure times of a few minutes allowed for a median signal-to-noise ratio (S/N) at 550 nm of 94.

\paragraph{UVES --} From September 2016 to August 2018, we used the UVES spectrograph \citep{dekker00} mounted on the UT2 telescope at the {\it Very Large Telescope} (Table~\ref{tab:list_spectro}) to carry out an analogous survey of 24 southern Cepheids. We used both UVES blue and red arms (centred at $\sim$437 and $\sim$760 nm, respectively) to expand our wavelength coverage. Given the fast acquisition rate, we acquired several consecutive (up to a dozen) spectra at each epoch and with each arm. We collected overall 981 and 1350 spectra with the blue and red arms of UVES, with a median S/N of 135 and 156, respectively.

\paragraph{HERMES --} From July 2014 to August 2018, we acquired 110 additional spectra with the HERMES spectrograph \citep[$R \sim$85~000,][]{raskin11} on the Flemish 1.2m at Roque de los Muchachos (La Palma) Observatory. This allowed us to complete our phase coverage of three targets \citep[V1334~Cyg, FF~Aql and W~Sgr, see][]{gallenne18b,gallenne18a}.

\subsubsection{Archive data}
For other spectrographs, we collected spectra that we already used in previous studies or that we retrieved from the ESO public archive\footnote{\url{http://archive.eso.org/wdb/wdb/adp/phase3_main/form}}.

\paragraph{HARPS --} We collected more than 400 archive spectra obtained with the {\it High-Accuracy Radial-velocity Planet Searcher} spectrograph mounted on the 3.6m telescope at La Silla observatory \citep[HARPS,][]{pepe02}. The HARPS spectrograph has the highest spectral resolution ($R \sim$115~000) and the best \vr~precision among all spectrographs currently implemented in our framework (Table~\ref{tab:list_spectro}). This highlights the importance of these data. We already presented most of these spectra in previous studies \citep[][]{nardetto06,nardetto07,nardetto09}. We retrieved the remaining spectra from the ESO database.

\paragraph{CORALIE --} We also gathered almost 500 spectra that we acquired with the CORALIE spectrograph \citep{queloz01a} at La Silla Observatory from 2013 to 2017. We already used some of these spectra in previous studies \citep[see \eg][]{gallenne18b}.

\paragraph{HARPS-North --} We used 103 high-resolution, high-S/N ($\gtrsim$~250) HARPS-North spectra of \dcep. The HARPS-North spectrograph is mounted on the TNG telescope at La Palma observatory \citep[see][and Table~\ref{tab:list_spectro}]{cosentino12}. We already presented these spectra in a previous study that highlighted their quality \citep{nardetto17}.

\paragraph{FEROS --} We finally retrieved almost one hundred FEROS spectra from the ESO public database. The FEROS instrument is a fibre-fed spectrograph mounted on the 2.2m MPG telescope at La Silla Observatory with a spectral resolution of $R \sim$48~000 \citep{kaufer98}. The FEROS data allowed us mostly to add three medium- to long-period Cepheids to our sample (UZ~Cet, AV~Sgr and V340~Ara, see Table~\ref{tab:sample}).

\section{Method}\label{sect:method}

\subsection{Principle: the CCF as the spectrum proxy}

Because of the physical characteristics of Cepheids, their location on the Hertzprung-Russell diagram, and their moderate rotation rates, Cepheid spectra typically exhibit hundreds or even thousands of narrow absorption lines. It is then possible to measure the Doppler shift (and hence, the \vr) for specific single lines of different depths \citep[see \eg][]{nardetto06}. However, single-line \vr~may differ from one line to another, due to velocity gradients or spectral peculiarities. The single-line \vr~precision also depends on the spectrum S/N, which might limit its use to high- or very high-S/N spectra \citep[typically with S/N above 75 to 100; see \eg][]{nardetto06,meunier17}.

An efficient approach to derive accurate \vr~that are representative of the full spectrum is to use a proxy for the spectrum instead of the spectrum itself. In other words, it consists in building a single global line profile that combines the useful information from all the spectral lines (Doppler shift, depth, width and asymmetry). This description corresponds well to the CCF. Building the CCF consists in cross-correlating the spectrum with a pre-defined template that is successively Doppler shifted \citep[][]{baranne79,queloz95}. This template can be a synthetic spectrum \citep[\eg~for the {\it Gaia Radial Velocity Spectrograph}, see][]{katz18}, a reference built from all spectra \citep{galland05}, or, more generally, an adapted binary template \citep[also named a binary mask, see \eg][]{queloz95,pepe02}. The binary designation refers to the fact that the template is equal to 1 (or $>$~0) at the wavelengths of the selected spectral lines and is equal to 0 everywhere else (spectrum continuum or rejected lines). Hence, the CCF includes the contribution of all the spectral lines included within the correlation template. It then provides a much higher S/N (and a better \vr~accuracy) compared to a single line \citep{pepe02,anderson17}. Alternative techniques or mathematical functions such as the spectral broadening function \citep[][]{rucinski92,rucinski99}, the auto-correlation function \citep{borra17}, or the least square deconvolution method \citep{britavskiy16}, have been proposed to characterise Cepheid line profiles. Still, the CCF remains both the easiest and most widely used method to study line profiles of Cepheids and other stellar pulsators \citep[see \eg][]{nardetto09,anderson17}. 

Specifically, the CCF built from a binary correlation template allows us to select which spectral lines to take into account in the global line profile and thus in the \vr~computation. This is a key factor to consider when trying to build homogeneous Cepheid \vr~time series. Given its relative simplicity, its widespread use, and its high interest in terms of line selection and template customisation, we thus decided to use CCF built from tailored correlation templates within our framework.

\begin{figure}[ht!]
\centering
\includegraphics[width=1.\hsize]{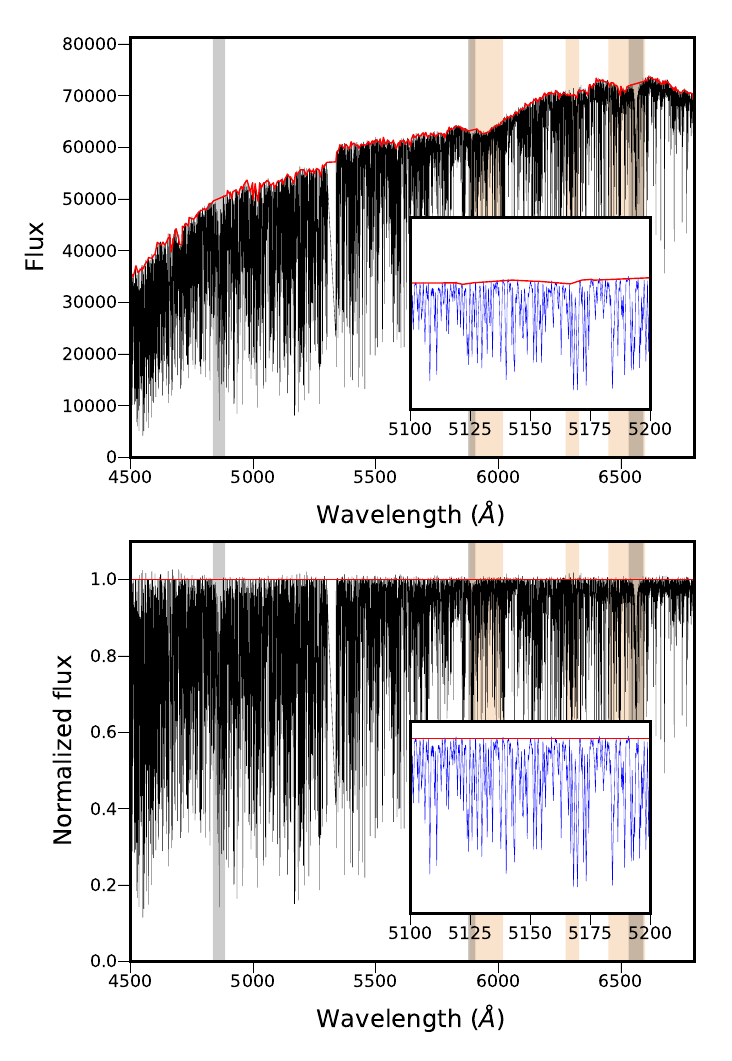}
\caption{Typical input observed spectrum. Top: input HARPS \bdor~1D spectrum (black solid line) on the \gr~wavelength range. The continuum interpolation is displayed in red. Broad deep lines that are excluded from the continuum interpolation are highlighted in grey, and wavelength ranges with strong tellurics are highlighted in orange. Bottom: normalised spectrum. The red solid line is normalised to unity. On both plots, the insert is a zoom on a 100-\AA~slice of the spectrum (in blue).}
\label{fig:sp}
\end{figure}

\subsection{Standardising the spectra}\label{method:spectra}

\begin{figure}[ht!]
\centering
\includegraphics[width=1\hsize]{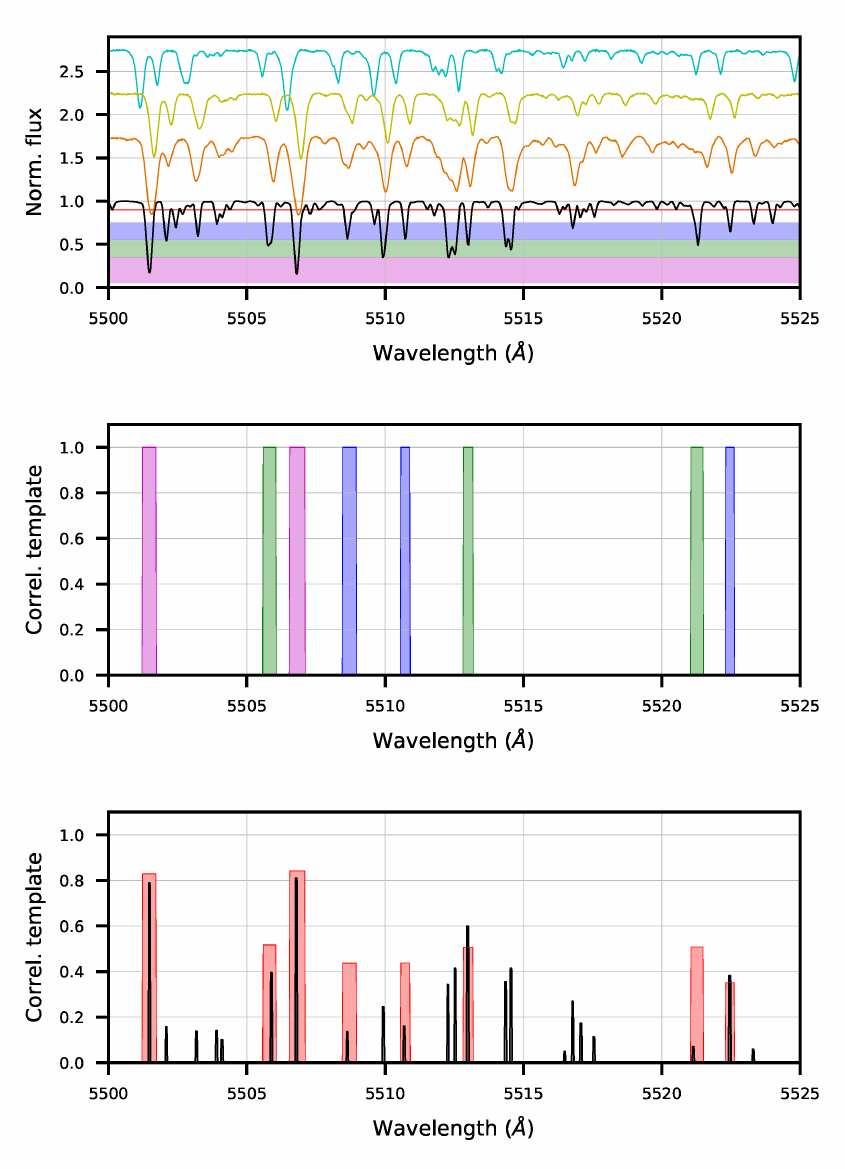}
\caption{Building tailored correlation templates. Top: reference synthetic \texttt{PHOENIX} spectrum (solid black line), with assumed limit on the continuum (solid red line). Line depth ranges corresponding to the \weak, \medm~and \deep~templates are highlighted (in blue, green and purple shades, respectively). Comparison with three observed Cepheid spectra, shifted in flux for clarity: \dcep~with HARPS-North ($\phi$~=~0.37, cyan line), \bdor~with HARPS ($\phi$~=~0.33, yellow line), and \lcar~with HARPS ($\phi$~=0.38, orange line). We point out that these spectra have not been Doppler-shifted from the stars' absolute velocities. Middle: corresponding \weak, \medm, and \deep~correlation templates (blue, green, and purple shades, respectively). Bottom: weighted \all~template (red shade). The default G2 template from the HARPS DRS (black shade) is added for comparison (see text).}
\label{fig:template}
\end{figure}

The main inputs of our framework are high-resolution spectra, pre-processed using the dedicated instrument pipelines. They are in a one-dimensional (1D) format (\ie~wavelength versus flux; typically the \texttt{s1d} format produced by ESO spectrograph data reduction systems or DRS), with the spectral orders already re-connected and re-sampled in wavelength. We chose the 1D format because it is a default product of the DRS of all the instruments included in this study, contrary to the 2D (\texttt{e2ds}) format which is not produced for UVES and FEROS to our knowledge. Furthermore, using the spectrum 1D format instead of the 2D format allows us not to have to correct for the instrumental response by spectral order (\ie~the blaze function), which would otherwise oblige us to introduce additional processing steps that might be less consistent from instrument to instrument. Given that classical Cepheid spectra typically exhibit thousands of narrow absorption lines, the order re-connection does not have any significant visual impact on the re-connected spectrum (Fig.~\ref{fig:sp}, top). The wavelength re-sampling typically leads to wavelength steps slightly smaller than the original CCD sampling (typically 0.01 - 0.02~\AA), meaning that the resulting 1D spectra are slightly over-sampled\footnote{see \eg~for SOPHIE: \url{http://www.obs-hp.fr/guide/sophie/reduction.html}}, which should not introduce any significant uncertainties. With the same view of achieving the greatest consistency of our CCF computation process, we consider that standardising our input spectra is a necessary step given the variety of the included spectrographs. First, we define three wavelength ranges ($\Delta\lambda$) on which the input spectra are cross-correlated. These ranges are fixed and defined to cover as much as possible the wavelength domains of the implemented spectrographs (Table~\ref{tab:list_spectro}):
\begin{enumerate}
\item A \gr~range with $\Delta\lambda$ = [4500-6800] \AA. This visible range corresponds to a wavelength domain covered by most of the implemented spectrographs. It also roughly corresponds to the wavelengths where spectra of classical Cepheids typically reach the highest flux density;
\item A \blue~range with $\Delta\lambda$ = [3900-4980] \AA. We defined this range to specifically cover the wavelength domain of the UVES blue arm;
\item A \red~range with $\Delta\lambda$ = [5700-8800] \AA. This range is covered by the UVES red arm, HERMES and FEROS.
\end{enumerate}
Second, we extract the continuum from each spectrum and normalise the spectrum following the same process for all spectrographs. Briefly, we build a continuum function or envelope by interpolating between a certain number of points that depends both on the spectrograph's resolution $R$ and on the absorption line density along the spectrum (Fig.~\ref{fig:sp}). Each interpolation point is carefully computed to be representative of the spectrum's continuum on the corresponding wavelength slice, by considering the flux value at 99\%~of the highest intensity over the considered slice while excluding possible cosmics. A similar normalisation process was used and validated by \cite{meunier17}. Broad deep lines (Hydrogen Balmer and Paschen series, as well as Calcium H and K lines) are excluded from the continuum function building. We simultaneously correct for the remaining cosmics, if any. The spectrum is then normalised by dividing it by this continuum function or envelope. Next, we correct the spectrum from the barycentric Earth radial velocity (or BERV), if necessary. Finally, we select the spectra to cross-correlate based on their S/N (taken at 5500 \AA~in the case of the \gr~wavelength range). We empirically put our S/N threshold to 30 to ensure a reasonably good CCF computation.

\subsection{Building tailored correlation templates}\label{method:template}

\renewcommand{\arraystretch}{1.25}
\begin{table*}[t!]
\caption{Main characteristics of our correlation templates. $T$ stands for the template transparency according to \cite{baranne79}, \ie~it represents the template $t(\lambda)$ transmission weighted by the covered wavelength range ($\Delta\lambda$): $T = 1/\Delta\lambda \times \int_{\Delta\lambda} t(\lambda) \ \mathrm{d}\lambda$. $N_{\rm \ell}$ stands for the number of lines included in the template and $\sigma_{\rm \ell}$ stands for the template mean line width (in \AA). The characteristics of the HARPS DRS G2 template considered over our \gr~range (last line) are showed for comparison (see text).}
\label{tab:template}
\begin{center}
\begin{tabular}{l l | c c c c c c r}\\
\hline
\hline
$\Delta\lambda$ & Template      & Line depth range & Mean depth  & $< \lambda >$ & $T$          & $N_{\rm \ell}$ & $\sigma_{\rm \ell}$     & Line weighting \\
                &               & (relative depth) & (rel. depth)& ($\AA$)       & (\%)         &            &  (\AA)             & (y/n)          \\
\hline
\hline
 \blue          & \medm         & [0.45 - 0.65]    &  0.55       &  4563         &  5.4         &  103       & 0.52              & n \\
\hline
 \gr            & \all          & [0.25 - 0.95]    &  0.63       &  5472         &  6.6         &  522       & 0.43              & y \\
 -              & \weak         & [0.25 - 0.45]    &  0.35       &  5827         &  2.6         &  146       & 0.35              & n \\
 -              & \medm         & [0.45 - 0.65]    &  0.56       &  5585         &  4.8         &  217       & 0.42              & n \\
 -              & \deep         & [0.65 - 0.95]    &  0.76       &  5280         &  4           &  163       & 0.51              & n \\
\hline
 \red           & \medm         & [0.45 - 0.65]    &  0.53       &  6852         &  2.7         &  88        & 0.48              & n \\
\hline
\hline
\gr             & G2/HARPS      &                  &  0.59       &  5339         &  1.8         &  1725      & 0.08              & y \\
\hline
\hline
\end{tabular}
\end{center}
\end{table*}
\renewcommand{\arraystretch}{1}

This step forms the focal point of our approach. Indeed, we use our custom correlation templates to cross-correlate all observed spectra on a given wavelength range regardless of the target and of the spectrograph. We illustrate our template building process in Fig.~\ref{fig:template} and we display the main properties of our templates in Table~\ref{tab:template}. To build our templates, we first generated a reference Cepheid synthetic spectrum over our three wavelength ranges. To do so, we used the radiative transfer \texttt{PHOENIX} code \citep{hauschildt99,hauschildt10}. \texttt{PHOENIX} is a non-Local Thermodynamic Equilibrium (nLTE) atmosphere model code, that uses spherically symmetric radiative transfer in the case of giant stars such as Cepheids \citep{hauschildt99b}. In terms of stellar physics, we adopted solar metallicities, which are suitable for Classical MW Cepheids with typical spectral types in the F8 to G5 range. We adopted a model with \teff~$= 5250$~K and \logg~$= 1$. This corresponds to a Cepheid somewhat colder and with a \logg~slightly smaller than the average of our sample \citep[\ie~it roughly corresponds to the average \teff~and \logg~of a $P$~$\sim$15- to $\sim$20-day Cepheid, see \eg][]{kovtyukh05}. We chose these \teff~and \logg~values as a compromise between obtaining a reference synthetic Cepheid spectrum with as much un-blended spectral lines of different depths as possible and staying close to the average properties of our sample. As a consequence, our adopted synthetic spectrum corresponds to a slightly later spectral type than the G2 templates classically used in the literature. We generated our synthetic reference spectrum over our three wavelength ranges with $R = 115000$, corresponding to our spectrograph with the highest resolution (HARPS, see Table~\ref{tab:list_spectro}).

From our reference spectrum, we selected the spectral lines to be included within our correlation templates following an approach similar to \cite{hindsley86}. First, we considered only lines stronger than 0.1 in relative depth (\ie~with a minimum normalised flux below 0.9), as an arbitrary limit between meaningful lines and the continuum of the spectrum. Then, we selected the lines to include based on their relative depth, in order to probe lines forming at different optical depths. We considered three line depth ranges (\ie~shallow, intermediate-depth and deep lines) and we built three corresponding templates on the \gr~wavelength range (\weak, \medm~and \deep~templates). We selected only un-blended lines separated from adjacent lines by more than 0.4~\AA~for deep and intermediate lines, and by more than 0.7 \AA~for shallow lines. Finally, we excluded any lines within wavelength ranges around broad non-metallic lines such as the Hydrogen Balmer and Paschen series \citep{anderson17}, as well as the Calcium II H and K lines and the Calcium II near-infrared triplet. We also excluded lines within wavelength ranges corresponding to strong telluric lines, based on inputs from the ESO \texttt{MOLECFIT} tool \citep{smette15} and on high-resolution spectra of Achernar (\ie~a fast-rotating Be dwarf star with very few and very broadened stellar lines). For each of these templates, the selected lines were uniformly put at a height of~1 within the template (\ie~no weighting according to the line depth, see Fig.~\ref{fig:template}, middle). Thus we roughly approximate the stellar atmosphere as a three-layer model, each layer being probed by a template. Having lines of same weight within each template means that the average line depth of the template is roughly equally representative of all the selected lines (Table~\ref{tab:template}). We tailored these three depth-specific templates (\weak, \medm, and \deep) to have a number of lines of the same order of magnitude, average wavelengths as close as possible and average line depths as different as possible to investigate the specific impact of the template average line depth on the \vr~(Sect.~\ref{res:line_depth}).

Next, we built a fourth template (\all) including all the lines selected within the three previous templates on the \gr~range. This time we weighted the template lines proportionally to their relative depth within the \texttt{PHOENIX} spectrum (Fig.~\ref{fig:template}, bottom). Such weighting is typical of the default templates used within the DRS \citep{pepe02}. We built this \all~template: {\it i}) to have reference \vr~time series based on a template with a higher number of lines (Sect.~\ref{res:catalogue}); {\it ii}) for comparison with default weighted DRS templates (Sect.~\ref{res:line_width}); and {\it iii}) to have a reference for the comparison of our depth-dependent templates (Sect.~\ref{res:line_depth}). Weighting the lines means that stronger lines have more weight (more impact) than weaker lines within the CCF and \vr~computation. Thus, the average line depth of the \all~template is 0.63 (weighted lines) instead of 0.58 without line weighting, while the other template characteristics (mean wavelength, mean line width) do not significantly change. This also allows us to have a fourth template more distinct in terms of average depth from the \medm~template (average depth $\sim$0.56, Table~\ref{tab:template}).

\begin{figure}[ht!]
\centering
\includegraphics[width=0.98\hsize]{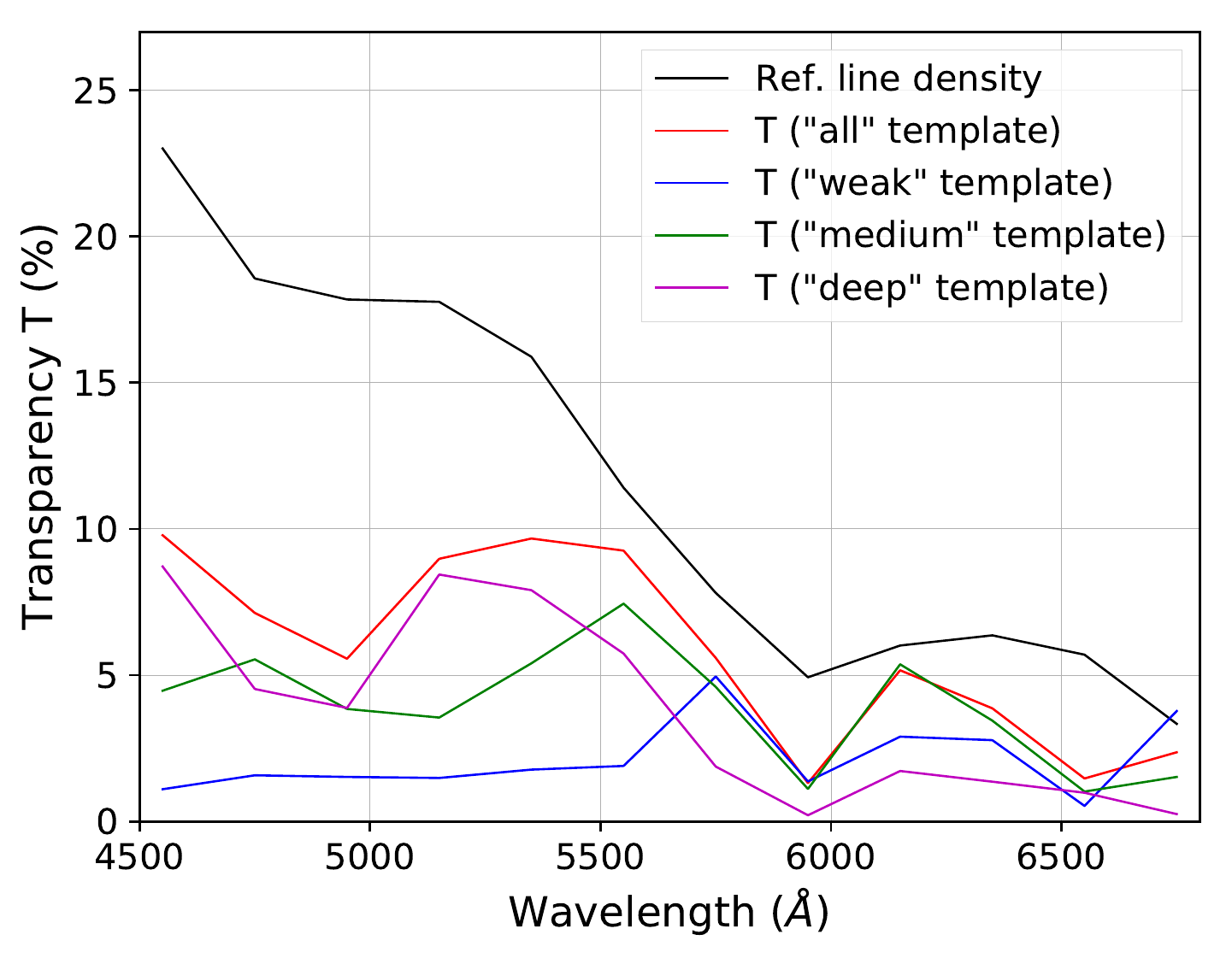}
\caption{Correlation template transparency $T$ (see Table~\ref{tab:template}) vs. wavelength for our \gr~$\lambda$~range. The line density of our reference spectrum is plotted in black and the transparency of our four correlation templates in red, blue, green, and purple (\all, \weak, \medm, and \deep~templates, respectively). The template transparency is equivalent to the selected line density within the template (see text).}
\label{fig:template_wave}
\end{figure}

Because of our strict spectral line selection, each of our three depth-specific templates includes $\sim$150 to $\sim$220 lines only. In contrast, the default DRS templates with a spectral type closest to classical Cepheids (\ie~typically G2-type templates adapted to Main-Sequence dwarfs) include thousands of often blended lines \citep{anderson16}. As an example, the HARPS DRS default G2 template considered over our \gr~range includes more than 1700 lines (see Table~\ref{tab:template} and Fig.~\ref{fig:template}, bottom). We also note that such DRS templates have very narrow lines (typical width $\sigma_{\ell} < 0.1$~\AA). In contrast, we fixed our template line width by considering the selected line width at 90\%~of the continuum within our \texttt{PHOENIX} synthetic Cepheid spectrum, \ie~average line widths larger by a factor 4 to 6 compared to the G2 template. We discuss in more details the rationale for this choice and its impact in Sect.~\ref{res:line_width}. We finally note that our correlation templates are sampled with a fine wavelength step of $\sim$0.02 \AA, meaning that our template lines typically cover 15 to 25 wavelength pixels.

\begin{figure*}[ht!]
\centering
\includegraphics[width=0.98\hsize]{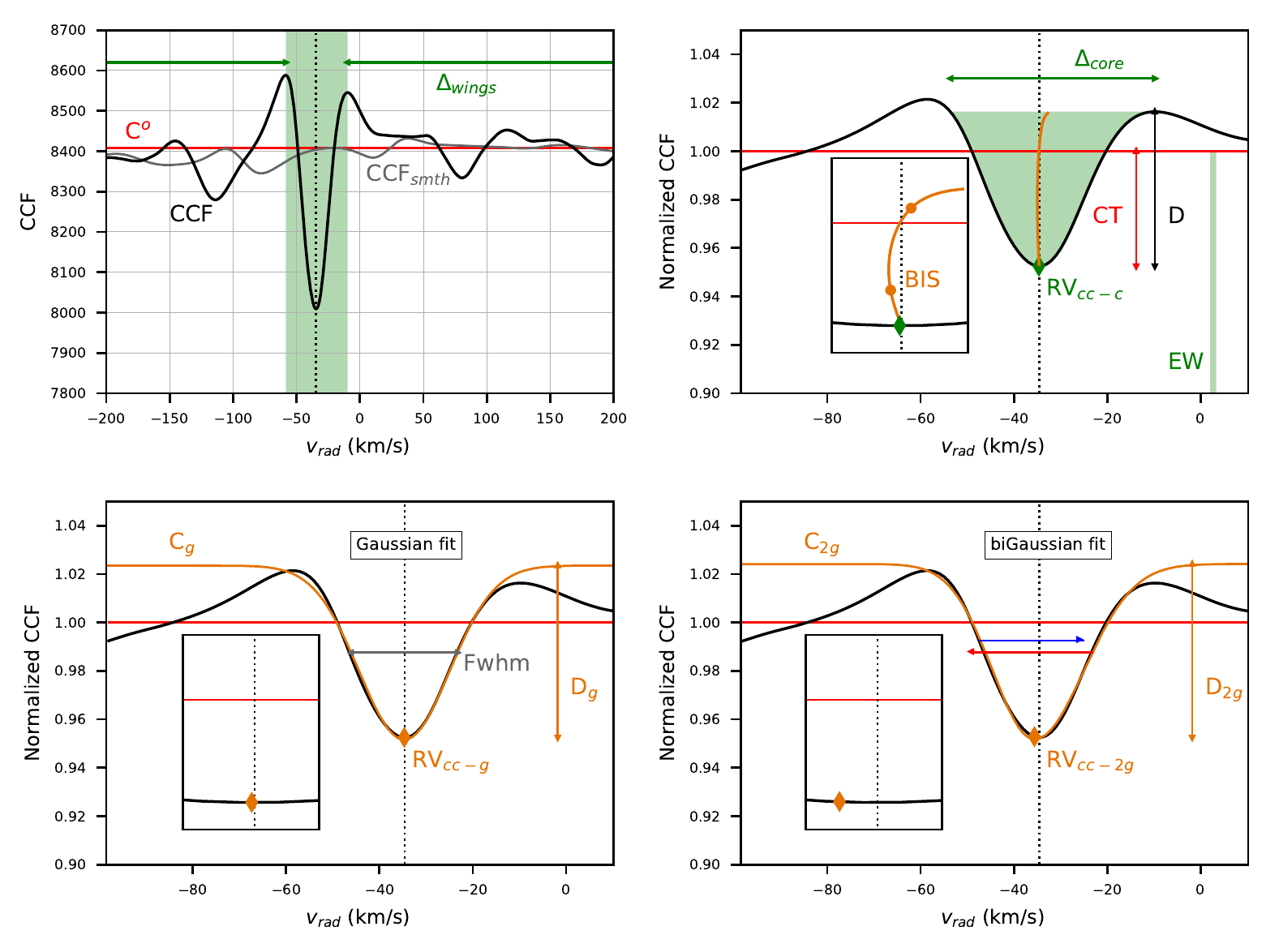}
\caption{CCF, main line profile observables and \vr~(based on a \dcep~HARPS-North spectrum cross-correlated with our \medm~template on the \gr~range, see text). From left to right and top to bottom: CCF, normalised CCF, normalised CCF Gaussian fit, normalised CCF biGaussian fit. On all plots the CCF is displayed as a black solid line, the CCF continuum as a solid straight red line and the \vr~at the CCF minimum as a vertical dotted black line. On the top left plot, the CCF core \vr~range (\ie~\dcore) is highlighted in green. On the top right plot, the area covered by the CCF core and used for the EW and \rvc~integration is highlighted in green. $CT$ and $D$ designate the contrast and depth of the CCF core, respectively (as defined in the text). On the bottom right plot, the FWHMs corresponding to the blue and red parts of our biGaussian model are displayed as a blue or red arrow, respectively. The inserts are zooms in \vr~on the CCF core.}
\label{fig:param_ccf}
\end{figure*}

The distribution of selected lines within our templates generally follows the spectral line density of our reference spectrum (\ie~the line distribution as a function of wavelength), with dips corresponding to telluric or broad line exclusion ranges (Fig.~\ref{fig:template_wave}). The slightly redder average wavelength of our \weak~template is induced by the increased line density (\ie~a higher line blending) in the bluer part of our reference spectrum. In addition, we built two other templates including intermediate-depth lines but covering this time our \blue~and \red~wavelength ranges. The number of included lines is relatively small, due to either the reduced wavelength span and strong line blending (in the case of the \blue~range) or the importance of the telluric ranges (in the case of the \red~range). However, our three \medm~templates have nearly similar average depths and average wavelengths separated by more than 1000 \AA~from each other (Table~\ref{tab:template}), to investigate the potential impact of the template wavelength range on the \vr~(Sect.~\ref{res:wave}).

\subsection{Characterising the CCF}

We cross-correlate each of our reduced spectra with our tailored templates, depending on the spectrograph and the covered wavelength range. We aim at extracting as much information as possible from the CCF, not only the selected Doppler shift or radial velocity. To do so, we use and derive different estimators of the shape of the CCF profile, that we discuss here. We specifically discuss the \vr~measurements later in Sect.~\ref{method:rv}. We give more technical details and formulae for the observable derivation in Appendix~\ref{appdix:method}. We illustrate our various observables in Fig.~\ref{fig:param_ccf} and summarise them in Table~\ref{tab:param}.

\renewcommand{\arraystretch}{1.25}
\begin{table*}[ht!]
\caption{CCF observables considered in this study. The line profile observable acronyms introduced here are used only for clarity in Fig.~\ref{fig:param_ccf} and in Appendix~\ref{appdix:method}.}
\label{tab:param}
\begin{center}
\begin{tabular}{l l l l l c | c}\\
\hline
\hline
CCF shape              & CCF wings          & CCF core        & CCF core & CCF core           & Doppler shift, & CCF quality \\
characteristics        &                    & depth           & width    & asymmetry          & \vr            &             \\
\hline
CCF-based proxies      & Continuum (\coo)   & Depth ($D$), EW      & EW       & BIS           & \rvc           & $Q$, \sncc  \\
-                      &                    & Contrast ($CT$)      &          &               &                &             \\
CCF Gaussian fit       & Offset (C$_{\rm g}$) & Depth ($D_{\rm g}$)   & FWHM     &               & \rvg           &             \\
CCF biGaussian fit     & Offset (C$_{\rm 2g}$)& Depth ($D_{\rm 2g}$)   &          & Asym. (\asym) & \rvbg &             \\
\hline
\hline
\end{tabular}
\end{center}
\end{table*}
\renewcommand{\arraystretch}{1}

\paragraph{CCF core, wings, and continuum --} Cepheid spectra may exhibit strongly asymmetric lines induced by the line-of-sight projection of the pulsation velocity. Given that we exclude most blended lines from our correlation templates (\ie~lines that usually smooth the CCF if taken into account), the typical shape of our CCFs may deviate significantly from that of a Gaussian \citep[Fig.~\ref{fig:param_ccf}, and see \eg][]{queloz95}. First, our CCFs typically exhibit two significant bumps or shoulders on both sides of the CCF core; second, the CCF shape outside of the CCF core is not completely flat. These effects are more noticeable for our CCFs than for CCFs computed with typical DRS templates because we use templates with a relatively small number of lines of variable depth, and because we reject most blended lines. Most often, studies that present CCF profiles show only the CCF core and not the CCF wings. We consider necessary to take into account both the core and the wings of the CCF for a proper CCF characterisation within our study. We compute our CCFs on an extended \vr~grid ranging from -200 to 200~\kms~in order to sample adequately the full CCF profile. The respective height of the two CCF shoulders depends on the CCF asymmetry and the direction of the Doppler shift, \ie~the CCF left shoulder is higher when the spectrum is blue-shifted (Fig.~\ref{fig:param_ccf}) and reciprocally. We define the CCF core as the area centred on the CCF main peak and below the CCF lower shoulder for practicality. Then the CCF wings include the whole \vr~ranges outside of the CCF shoulders, and the CCF pseudo-continuum (\coo) is defined as the average value of the CCF wings. The CCF can be normalised by dividing it by the value of \coo, as done for the input spectra.

\paragraph{Modelling the CCF --} We derive some of our line profile observables through fitting the CCF core by parameterised models. We use here both a classical four-parameter Gaussian model (offset, depth, width and Doppler shift) and a biGaussian model (offset, depth, asymmetry and Doppler shift) where we distinguish between the blue and red parts of the CCF core (see more details in Appendix~\ref{appdix:method}).

\paragraph{CCF depth and width --} We define the CCF core depth ($D$) as the difference between the maximum and the minimum of the CCF core, that we either measure directly on the CCF or through the Gaussian or biGaussian models. We distinguish this CCF depth from another separate observable that we name here the CCF contrast ($CT$) and that we define as the difference between our CCF continuum and the minimum of the CCF core (Fig.~\ref{fig:param_ccf}, top right). Our main observable for the CCF width is the Full Width at Half Maximum (FWHM) of the CCF core Gaussian model (Fig.~\ref{fig:param_ccf}, bottom left). Finally, we compute the Equivalent Width (EW) of the CCF core in a similar way as what is commonly done for single spectral lines \citep[see \eg][]{kovtyukh05}. The EW is a mixed proxy of the CCF core depth and width. The variability of the depth and width of the CCF profile is directly related to the various quantities that have a broadening effect on the spectral lines (\ie~the pulsation, but also and mostly the effective temperature \teff, the turbulence and the Cepheid rotation rate). For example, the FWHM has been used as an estimator of the micro-turbulence velocity \citep{borra17}.

\paragraph{CCF asymmetry --} We derive the bisector of the CCF core as a classical way to estimate its asymmetry. Our proxy is the Bisector Inverse Span \citep[BIS,][]{queloz01b}, \ie~the \vr~difference between the top and the bottom of the CCF core bisector (Appendix~\ref{appdix:method}). According to authors \citep{anderson16,britavskiy17,anderson19}, the BIS is a good estimator of the line profile asymmetry of stellar pulsators such as Cepheids. Another asymmetry proxy is the line asymmetry estimator defined by \cite{nardetto06} based on the biGaussian model of single spectral lines. By analogy, we derive a similar asymmetry proxy from our CCF biGaussian model through the comparison of the width of the blue and red parts of the CCF core (Fig.~\ref{fig:param_ccf}, bottom right, and see Appendix~\ref{appdix:method}).

\begin{figure*}[ht!]
\centering
\includegraphics[width=1.\hsize]{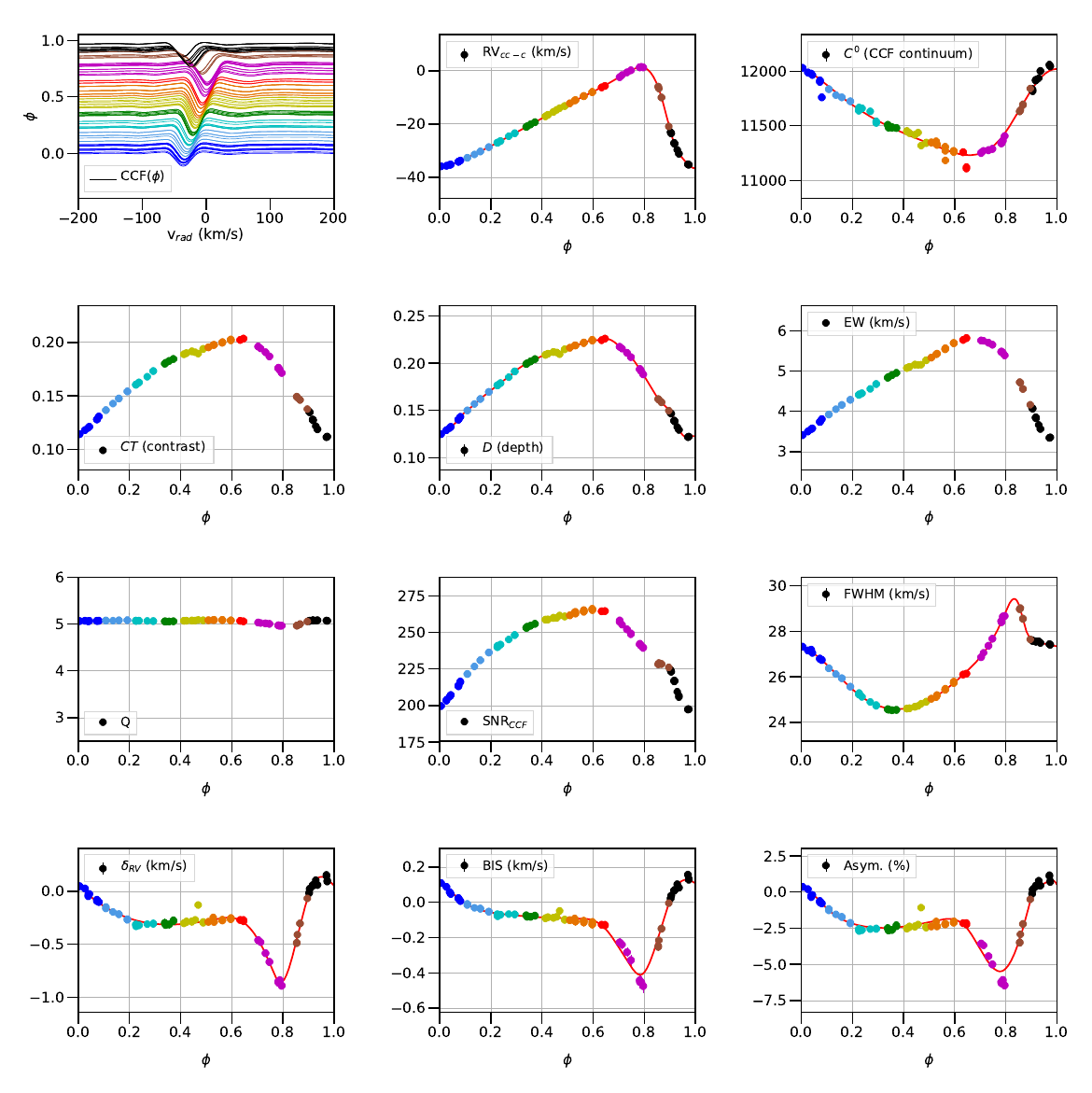}
\caption{Normalised \dcep~CCFs (top left plot) and main observables phased along the pulsation period, colour-coded with the phase $\phi$. Based on \dcep~HARPS-North spectra cross-correlated with our \all~template on our \gr~range. On the bottom left plot, $\delta_{RV}$ denotes the difference between our Gaussian (\rvg) and biGaussian (\rvbg) \vr. Measurements are displayed as coloured dots and interpolated spline curves as red solid lines, if any.}
\label{fig:catalogue}
\end{figure*}

\paragraph{CCF quality --} Depending on the number and the strength of the lines selected within, the correlation template has a direct impact on the resulting CCF, its global shape and its depth. In particular, selecting lines of reduced strength (\ie~shallower lines) reduces the amount of signal contained by the CCF core with respect to the CCF wings. Thus, the reliability and accuracy of the derived \vr~and line profile observables will be impacted. Here, we find it necessary to assess the quality of our derived CCF, in order to figure how much confidence we can put on our observables. We define the two following criteria (see formulae in Appendix~\ref{appdix:method}):
\begin{enumerate}
\item a CCF quality factor $Q$, defined as the ratio of the CCF contrast $CT$ over the CCF short-range \vr~variability, \ie~the standard deviation of the difference between the original CCF and the CCF smoothed over a given \vr~window (CCF$_{smth}$ in Fig.~\ref{fig:param_ccf} top left);
\item a CCF signal-to-noise ratio (\sncc), defined as the ratio of the CCF core depth $D$ over the standard deviation of the CCF wings.
\end{enumerate}
The $Q$ criterion estimates how much noisy or dispersed is the whole CCF and how well the CCF core can be distinguished from the CCF wings. This is important with respect to the convergence and reliability of our CCF Gaussian or biGaussian models and our observable automatic computation. In the following, we adopt an arbitrary minimal threshold of $Q = 4$ for good-quality CCFs (more details in Sect.~\ref{sect:results}). \cite{baranne79} also defined a CCF quality factor, but it was directly dependent on the spectrum exposure time and its photon noise. Here our $Q$ criterion is purely CCF-specific. Defining such a quality criterion helped us to build our line depth-specific correlation templates. Our second criterion \sncc~estimates the amount of signal within the CCF core, weighted by the CCF wing global dispersion. We used \sncc~to derive the uncertainties on some of our CCF-based proxies (see Appendix~\ref{appdix:method}).

\subsection{Computing the radial velocities}\label{method:rv}

\begin{figure}[ht!]
\centering
\includegraphics[width=1.\hsize]{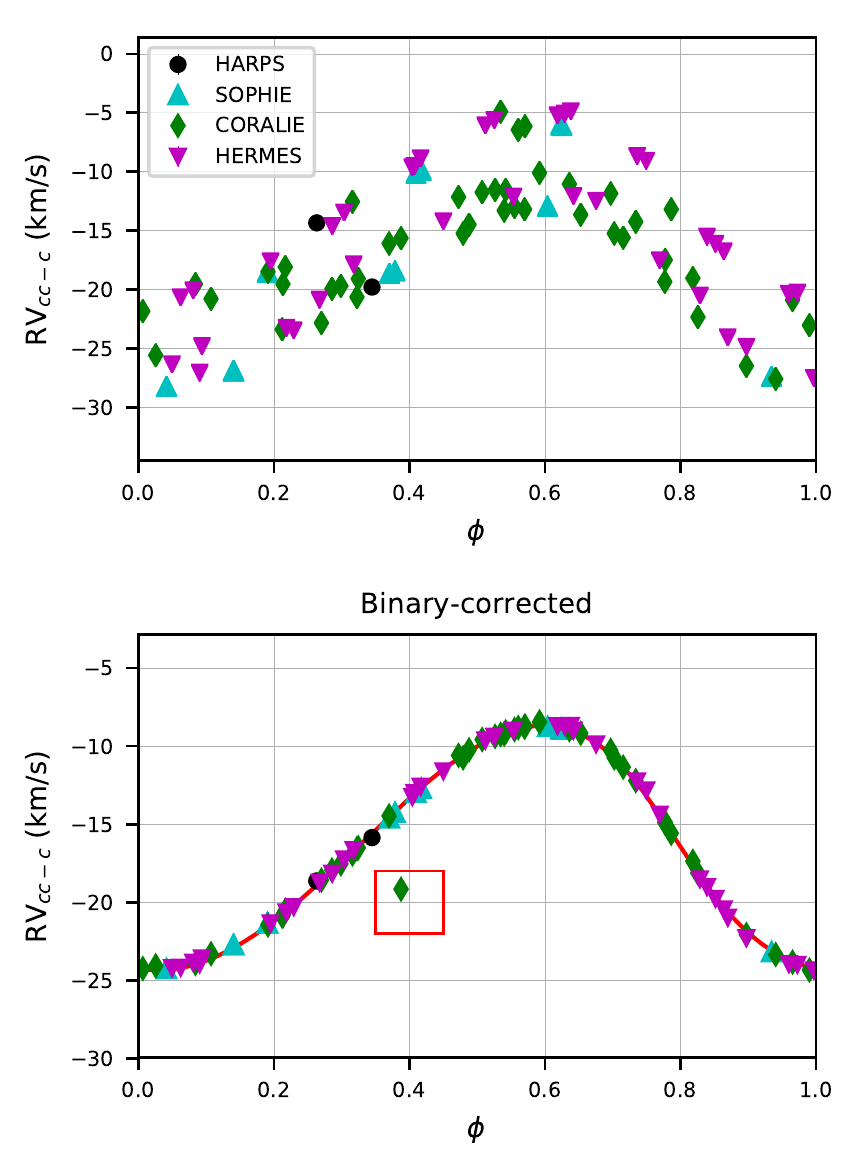}
\caption{Combined \vr~of FF~Aql. Top: centroid \vr~(\rvc) obtained with the \all~template on the \gr~range for four spectrographs, phased ($\phi$) along the Cepheid pulsation period and not binary-corrected. Bottom: the same, but the \vr~have been corrected from the Keplerian orbit of the binary companion. Our best spline curve is displayed as a red solid line. The \vr~are corrected only from the Keplerian orbit (no offset correction), based on the orbital parameters recently computed by \cite{gallenne18b}. In the red box is one of our few CORALIE \vr~outliers (see text), that we did not take into account to build the spline curve.}
\label{fig:binary}
\end{figure}

We emphasise here again that any Cepheid (or pulsating star) \vr~measurement is somewhat biased with respect to the pulsation-induced asymmetry of the line or CCF profile. We decided here to merely implement three different (and well-known) ways to compute \vr~measurements. We did not try to definitely assess which method is to be preferred. Generally speaking, we consider that several different \vr~computation methods should always be considered for Cepheids and pulsating stars \citep{burki82}.

\paragraph{Centroid \vr~--} The CCF Doppler shift can be quantified by computing the centroid or barycentric velocity (hereafter \rvc, Appendix~\ref{appdix:method}). In the same way as done by \cite{nardetto06} for single spectral lines, the CCF \rvc~corresponds to the first moment of the CCF core profile \citep[see ][and Fig.~\ref{fig:param_ccf}, top right]{hindsley86}. Studies of the Cepheid $p$-factor decomposition have favoured single-line centroid \vr~compared to other single-line \vr~computation methods as they are independent from rotational and turbulent broadening \citep{burki82,nardetto06}. However, they require a high enough S/N. 

\begin{figure}[ht!]
\centering
\includegraphics[width=1.\hsize]{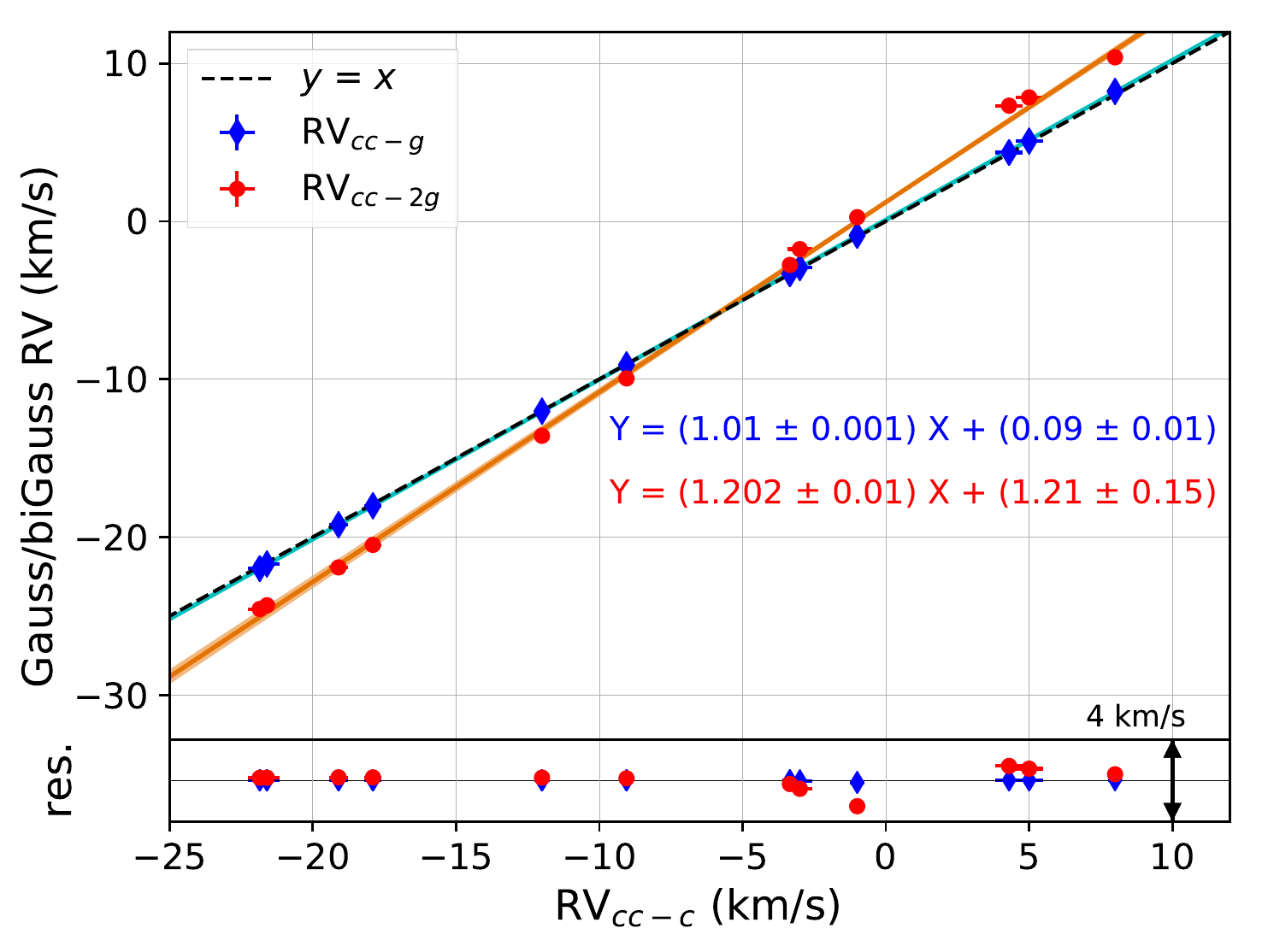}
\caption{Comparison of \vr~computation methods for FM~Aql SOPHIE \vr~(\gr~range, \medm~template). The $x$-axis corresponds to \rvc. \rvg~and \rvbg~are displayed as blue diamonds and red dots, respectively. We point out that the uncertainties on all \vr~are displayed but are not necessarily visible. The best linear regressions are displayed as cyan and orange solid lines (along with their 1$\sigma$ uncertainty in same-colour shades), respectively. The black dashed line corresponds to a slope of 1. The bottom insert corresponds to the residuals of the two linear regressions (same colour code).}
\label{fig:rv_method_exp}
\end{figure}

\paragraph{Gaussian \vr~--} The most classical and most widely used way to compute the \vr~is to fit the CCF profile with a Gaussian model (Fig.~\ref{fig:param_ccf}, bottom left). The derived Gaussian radial velocity (hereafter \rvg) is less sensitive to scatter in the spectrum than the CCF first moment, and is thus more stable and less dependent on the S/N of the spectrum \citep{anderson17}. However, \rvg~is potentially biased for Cepheids and other stellar pulsators as it accounts badly for the CCF profile asymmetry at high pulsation velocities \citep{nardetto06}.

\paragraph{biGaussian \vr~--} A solution to reproduce more closely the CCF asymmetry is to fit the CCF core with a biGaussian model instead of a simple Gaussian \citep[as first done by][for single-line \vr~computation]{nardetto06}, \ie~by fitting separately the blue and right parts of the CCF core profile (Fig.~\ref{fig:param_ccf}, bottom right). This gives us a third, biGaussian, \vr~value (hereafter \rvbg).

\section{Results}\label{sect:results}

\begin{figure*}[ht!]
\centering
\includegraphics[width=0.95\hsize]{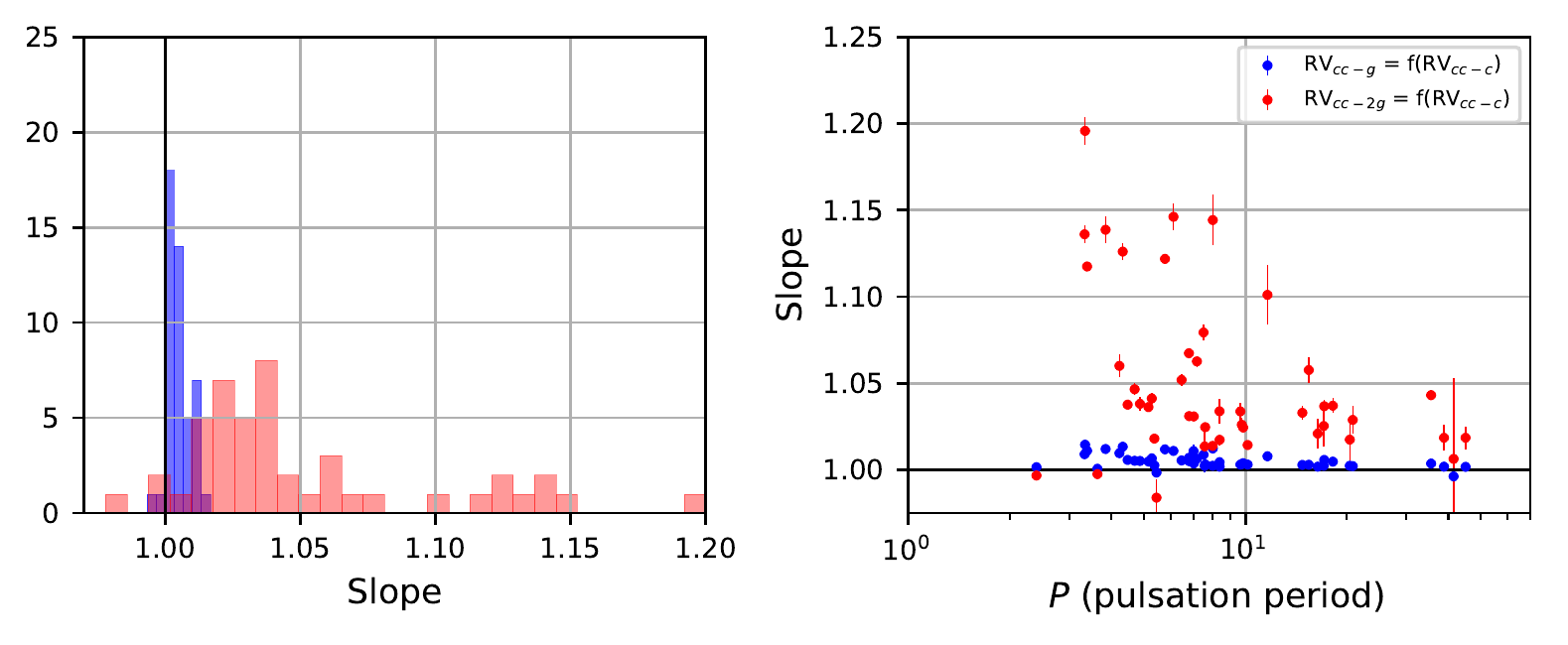}
\caption{Left: slope distribution for the \rvg~vs~\rvc~and \rvbg~vs~\rvc~linear regressions (in blue and red shades, respectively). Right, the sames slope values versus the pulsation period for each target. Computation made based on our \all~template on our \gr~range.}
\label{fig:rv_method}
\end{figure*}

\subsection{A consistent catalogue}\label{res:catalogue}

We computed the CCFs, corresponding line profile observable and quality proxy time series, and corresponding \vr~time series for our whole 64-Cepheid sample, using our six correlation templates and the input spectra standardised on our three wavelength ranges. This makes up a large, homogeneous catalogue of MW Cepheid CCF and \vr. The full catalogue is to be published on-line, \ie~the CCFs, various time series as well as our correlation templates. We display a (small) example of our data (\eg~CCFs, \vr~and line profile observables deduced from \dcep~HARPS-North spectra cross-correlated on the \gr~range with the \all~template) in Fig.~\ref{fig:catalogue}. The displayed data look robust and behave as expected along the pulsation phase. The CCF asymmetry proxies (BIS and biGaussian asymmetry) show the same behaviour and are correlated to the difference between biGaussian and Gaussian \vr~\citep[in agreement with the results of][]{anderson16}. In this example, our CCF quality factor $Q$ is high and nearly constant at all phases, which we will show latter to be a sign of good CCF quality (Sect.~\ref{sect:results}). Our CCF \sncc~behaves in correlation with the CCF depth, as expected: it is the highest when the CCF is the deepest, \ie~when the Cepheid reaches it largest radius. On the opposite, the CCF FWHM is the largest at the end of the contraction phase \citep{nardetto06}.

We provide the \vr~time series as processed, \ie~with the corresponding observation Modified Julian Day (MJD) but without any correction done on the \vr, except for the correction from the BERV done on the input spectra themselves if necessary (Sect.~\ref{method:spectra}). Thus we provide the \vr~without correcting for a potential binary companion and without removing the absolute star \vr~(or systemic velocity). For some of our targets (most of them being also spectroscopic binaries, see below), we have enough data to sample the pulsation phase adequately with several spectrographs. However, the data are not enough for us to clearly estimate instrumental \vr~offsets from spectrograph to spectrograph because such offsets are typically of the order of 100~\ms~or below, \ie~very small compared to the Cepheid \vr~variability (pulsation- and binary-induced). Some of these instrumental spectrograph-to-spectrograph \vr~offsets have been previously measured with a great accuracy \citep{soubiran13,gallenne18a}. In terms of outliers, we have only to note a very few \vr~problematic data (\ie~that exhibit unexpected offsets for all templates and all methods). Those concern a few CORALIE spectra acquired between MJD~57196 and MJD~57206 (\ie~in mid-June 2015, see Fig.~\ref{fig:binary}, bottom plot).

\subsection{Spectroscopic binaries}\label{res:binaries}

Most of MW Classical Cepheids have been revealed to be components of binary or even multiple systems \citep{kervella18}. Here, we are mainly concerned with single-lined spectroscopic binaries (SB1s), \ie~for which the companion signature is noticeable in the derived primary Cepheid \vr. We detected a number of 18 unambiguous SB1s within our Cepheid sample, with 11 other Cepheids exhibiting \vr~scatter that hint towards a companion (Table~\ref{tab:sample}). For some of these binaries, new derivations of the companion orbital parameters were recently performed based on the data presented here in published studies \citep[see][and Fig.~\ref{fig:binary}]{gallenne18b,gallenne18a}. We redirect the interested reader towards the latter studies, since Cepheid SB1s are not the focus of the present study. Nonetheless, we propose in Appendix~\ref{appdix:bin} a new estimation of the companion orbital parameters for two of our SB1 targets (SU~Cyg and V496~Aql), to highlight the interest of our new \vr~data in combination with previous \vr~data. In the following, we did not find it necessary to correct the \vr~data for the companion signatures, as we mainly compare the same \vr~time series computed in different ways.

\subsection{\vr~computation method}\label{res:method}

Here, we investigate the specific impact of the \vr~computation method on the Cepheid \vr~measurements. We compare the three methods that we implemented here, \ie~the centroid, Gaussian and biGaussian \vr~(\rvc, \rvg, and \rvbg, respectively). For this comparison, we limit ourselves to \vr~measurements made with a given correlation template on the \gr~range. We select Cepheids with a good sampling of the pulsation curve within these constraints only, \ie~47 targets. For each of these targets, we compute two linear regressions: (1) of \rvg~versus \rvc~measurements and (2) of \rvbg~versus \rvc~measurements (Fig.~\ref{fig:rv_method_exp}). We then study the distribution of the values of the slope of these regressions. We consider that such an approach is safer and more reliable than comparing the peak-to-peak amplitudes of the two \vr~time series phased along the Cepheid pulsation period \citep[as done by \eg][]{nardetto07}. First, it allows to take into account all the \vr~measurements (and not only the extremal \vr~values). Second, it should make the comparison of the two \vr~time series less prone to biases induced by \eg~shockwaves \citep{nardetto18b} or cycle-to-cycle \vr~variability \citep{anderson14,anderson16}, that affect especially the extremal \vr~measurements. We use the same approach in the following (Sects.~\ref{res:wave}, \ref{res:line_width}, \ref{res:line_depth}).

Overall, we find Gaussian \vr~to have slightly larger amplitudes than centroid \vr, and biGaussian \vr~to have significantly larger amplitudes than both other methods. For the \all~template, we estimate \rvg~to be larger than \rvc~by $\sim$1\%~overall, and \rvbg~to be larger than \rvc~by $\sim$3-4\%~overall (Fig.~\ref{fig:rv_method}, left). This agrees with biGaussian \vr~being more sensitive to the CCF asymmetry, as found in previous studies \citep{nardetto06,anderson16}. This trend seems to be more pronounced:
\begin{enumerate}
\item for shallower lines: with our intermediate-line template, we find \rvg~to be $\sim$1-2\%~larger than \rvc~and \rvbg~to be $\sim$5\%~larger than \rvc~overall;
\item for shorter pulsation periods (Fig.~\ref{fig:rv_method}, right). 
\end{enumerate}
Both results agree with shallower lines and corresponding CCFs being more asymmetric than deeper ones \citep{anderson16}. We also note that the dispersion of the linear regression slope values is much higher for biGaussian \vr~than for Gaussian \vr. The Gaussian model is probably more robust than the biGaussian model because it has one less fitting parameter. If confirmed, it would also mean that a robust $p$-factor distribution over a large Cepheid sample would be more difficult to obtain with biGaussian \vr~than with centroid or Gaussian \vr. To conclude here, we show and confirm the significant impact of the \vr~computation method on the deduced CCF \vr~time series. 

\subsection{Template wavelength range}\label{res:wave}

Here, we investigate the impact of the wavelength range ($\Delta\lambda$) on which we compute our CCFs on the \vr~measurements. To do so, we use the \vr~measurements that we obtained based on our three \medm~templates spanning the three pre-defined wavelength ranges (\blue, \gr, and \red). First, to compare \gr~and \red~CCFs and \vr, we use our three targets observed extensively with FEROS (UZ~Sct, V340~Ara and AV~Sgr), as well as our three HERMES targets (V1334~Cyg, FF~Aql and W~Sgr). For these instruments and targets, the same acquired spectra cover both the \gr~and \red~wavelength ranges, which allows us to make a direct comparison between the corresponding \gr~and \red~data. Second, to compare our \blue~and \red~wavelength ranges, we use our UVES data (24 targets in all, Table~\ref{tab:sample}), as the corresponding \blue~and \red~spectra were acquired at the same observation epochs (hence allowing for a direct comparison). For each target and each UVES arm, we average the successive CCFs obtained at each epoch of observation (see Sect.~\ref{sect:survey}) and we derive the corresponding observables, in order to have the same number of measurements for the blue and red arms.

\begin{figure*}[ht!]
\centering
\includegraphics[width=1\hsize]{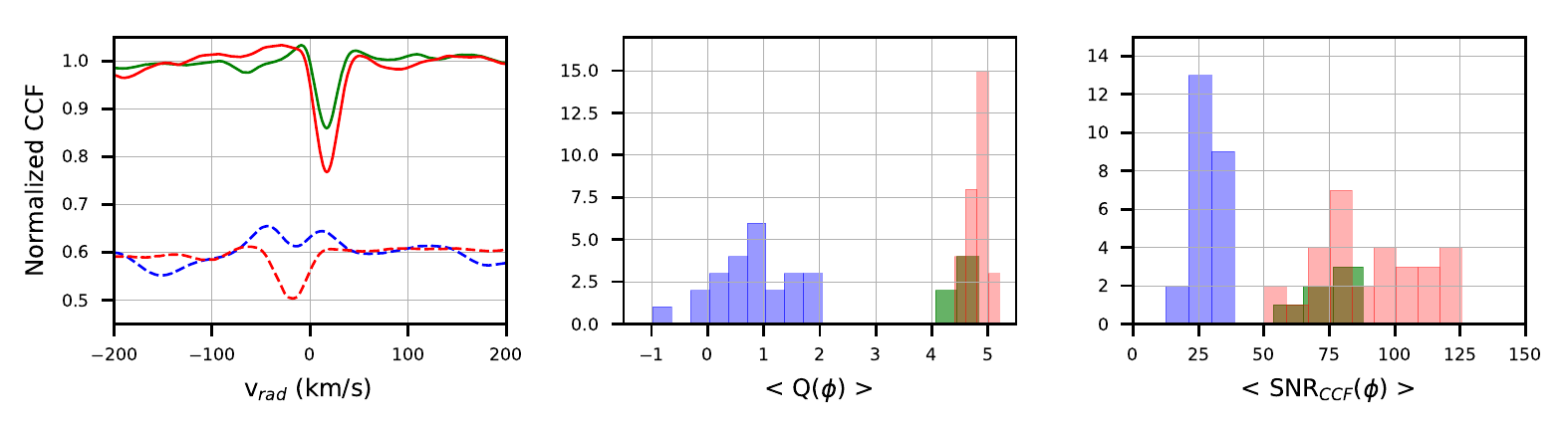}
\caption{CCF quality vs. wavelength range. Left (top, solid lines): comparison of a \gr~and a \red~CCF of UZ~Sct based on the same FEROS spectrum; comparison of a \blue~and a \red~CCF of BG~Cru (bottom; dashed lines) based on UVES spectra acquired at the same observation epoch with the blue and red arms of UVES, respectively. The bottom UVES CCFs have been shifted downwards for clarity. Middle: histogram of the CCF quality factor $Q$ averaged over the pulsation period $P$ for each studied Cepheid (green: six FEROS and HERMES targets; blue: 24 UVES blue arm targets; red: 30 FEROS, HERMES, and UVES red arm targets; see text). Right: the same, for the histogram of our \sncc~proxy averaged over $P$ for each target.}
\label{fig:qual_wavrange}
\end{figure*}

We look first at the general CCF quality for each wavelength range (Fig.~\ref{fig:qual_wavrange}). We note that our \gr~and \red~(intermediate-line template) CCFs exhibit a nearly constant average quality factor $< Q >$ (around 4.5-5). On the contrary, our \blue~CCFs show a decreased $< Q >$ that is much more variable (between -1 and 2, Fig.~\ref{fig:qual_wavrange} middle plot). The \gr~and \red~CCFs exhibit a well-defined and relatively deep core, while the \blue~CCFs are much more noisy and exhibit a shallower core (Fig.~\ref{fig:qual_wavrange}, left plot). Overall, we note that our good-quality CCFs have a nearly constant $Q$ above $\sim$4 (see also Fig.~\ref{fig:catalogue}) independently of their depth. On the contrary, when the ratio between the contrast of the CCF core and the dispersion of the CCF continuum is decreased enough, the CCF $Q$ factor starts to decrease (below 4) and becomes significantly variable. This led us to define a CCF quality threshold of $Q = 4$. When looking at our other proxy \sncc, it is more variable: \red~CCFs exhibit the highest \sncc~values (between 70 and 130 for intermediate-depth lines, Fig.~\ref{fig:qual_wavrange} right plot), while \gr~CCFs have somewhat lower \sncc~values (in agreement with the respective CCF depths). Finally, \blue~CCFs show understandably much smaller \sncc~values (below 40).

\begin{figure*}[ht!]
\centering
\includegraphics[width=1\hsize]{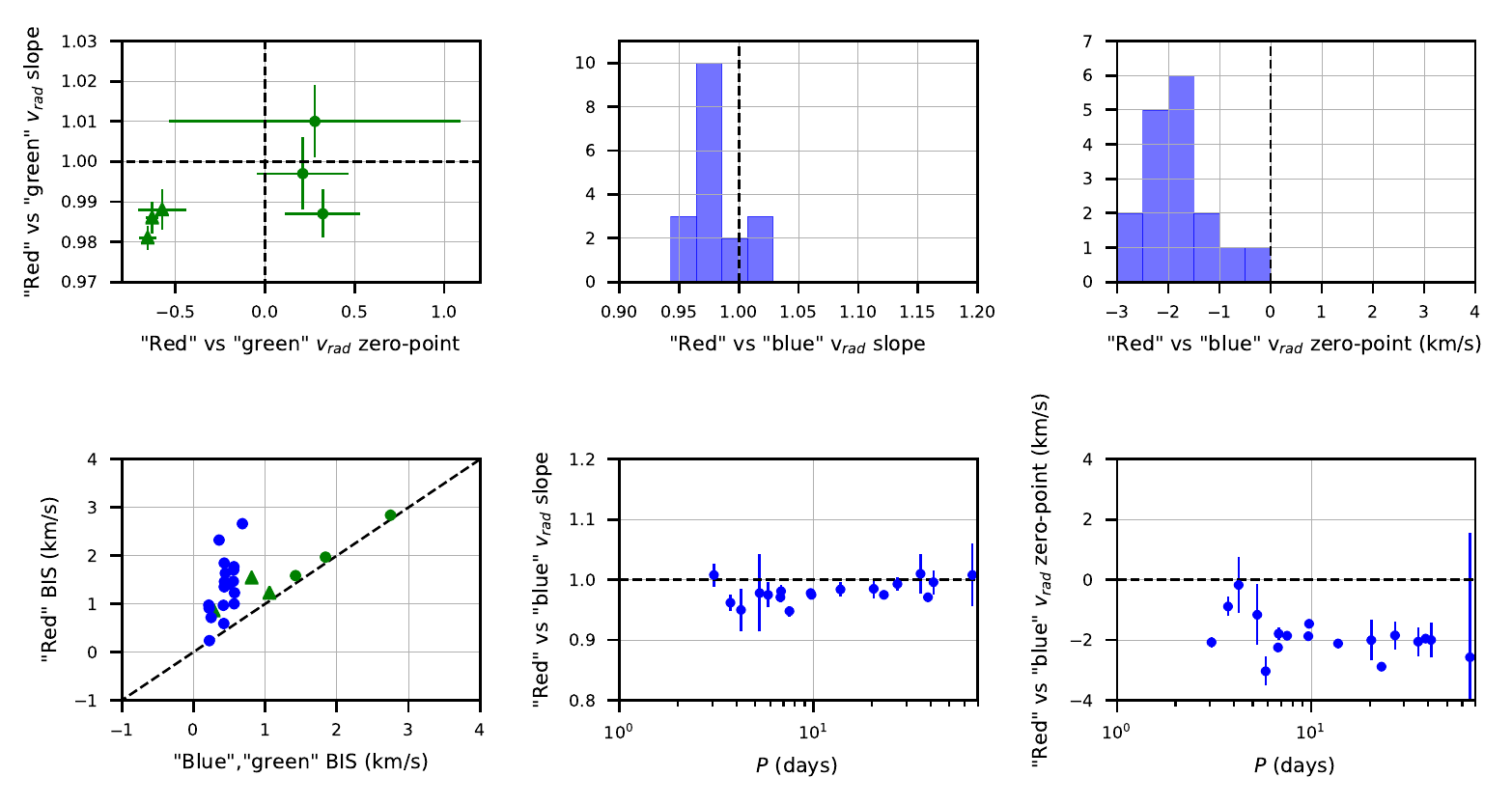}
\caption{CCF \vr~vs. wavelength range. Top: \red~vs. \gr~linear regression for our six targets (see text): the regression slope is plotted vs. the regression zero-point with green diamonds for HERMES targets and green dots for FEROS targets (left). Histogram of \red~vs. \blue~\vr~slope for our UVES targets (middle). The same, for the \vr~zero-point (right). Bottom: amplitude of variation over $P$ of the CCF BIS, plotted for our \red~range vs. our \blue~or \gr~range, for the same targets as above (left). Distribution of the \red~vs. \blue~\vr~slope with $P$ for our UVES targets (middle). The same, for the \vr~regression zero-point (right).}
\label{fig:vrad_wavrange}
\end{figure*}

This quality difference between \blue~and \red~UVES CCFs does not originate in the number of lines included within the respective correlation templates (Table~\ref{tab:param}). Rather, it originates in the much increased spectral line density on the \blue~range compared to the \red~range. Between 3900 and $\sim$5000~\AA, typical Cepheid spectra are crowded with spectral lines that are often blended with each other. Even if we tried to select only un-blended lines from our synthetic Cepheid spectrum, this line density has nonetheless an impact on our CCFs. Meanwhile, there are much less spectral lines on the \red~range, that are much more separated from each other. This explains the increased quality of the \red~CCFs.

Second, we look at the \blue~and \red~\vr~themselves. For each of the selected targets, we compute the linear regression of the \red~versus \gr~\vr~(FEROS and HERMES targets) or \red~versus \blue~\vr~(UVES targets). Overall, it is difficult to detect a definitive trend. Yet, the average slope of the linear regression seems to be slightly below 1 ($\sim$0.97-0.98, see Fig.~\ref{fig:vrad_wavrange}). If confirmed, this would mean that Cepheid \vr~amplitudes of variation decrease at larger wavelengths. This would agree with the findings of \cite{nardetto09}, who reported a linear decrease of the Cepheid \vr~peak-to-peak amplitude, based on CCFs computed order-by-order with the HARPS DRS and its classical G2 template. Yet, we do not find it clear whether this trend depends on the Cepheid pulsation period. We emphasise that our result remains to be confirmed given the relatively few number of \vr~measurements used to perform the linear regression for each target. In the context of this study, we also consider this wavelength effect (between our different $\lambda$ ranges) to be small enough to be neglected within a given $\lambda$ range. Such studies would need to be extended to infrared (IR) wavelengths in order to be confirmed. \cite{nardetto18} did not find a significant difference between the \vr~curve amplitude of \lcar~measured in the optical and on an IR line, respectively.

If considering the zero-point, there is no clear result for our \gr~versus \red~\vr~regressions given that the zero-point value seems to depend on the spectrograph (Fig.~\ref{fig:vrad_wavrange}, top left). When looking at our UVES targets, there seems to be a consistent offset of $\sim$1.5-2 \kms~between the \vr~measured with the blue and red arms of UVES. It is not clear whether this is related to the wavelength or to technical differences in the acquisition of the UVES blue and red spectra \citep{molaro08}. Finally, we find the BIS amplitude over the Cepheid pulsation period to be slightly increased over our \red~range compared to our \blue~or \gr~ranges (Fig.~\ref{fig:vrad_wavrange}, bottom right).

\begin{figure*}[ht!]
\centering
\includegraphics[width=\hsize]{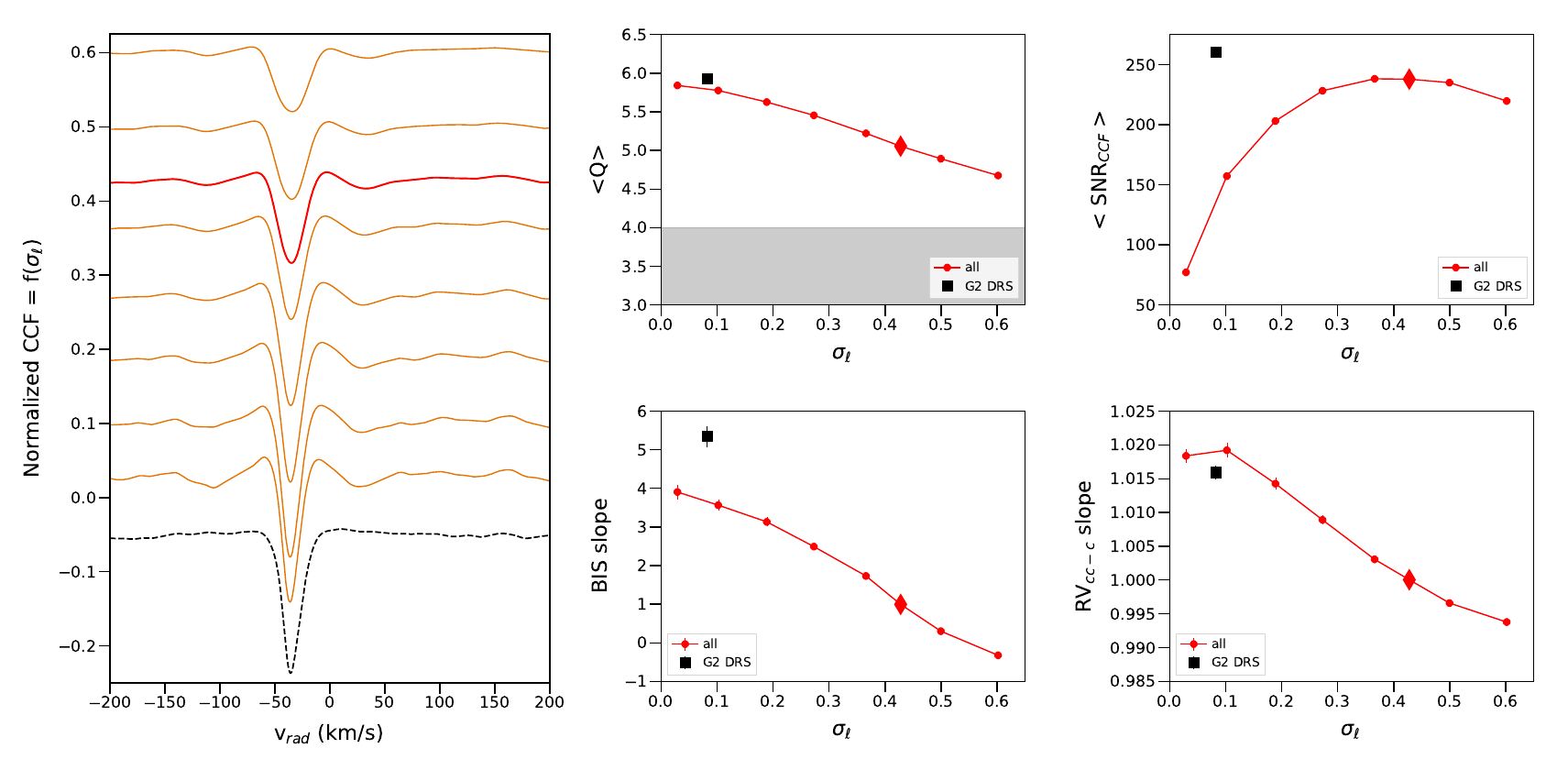}
\caption{CCF and \vr~vs. template average line width $\sigma_{\ell}$. On the left are displayed the CCFs resulting from the cross-correlation of one observed \dcep~spectrum (see text) with the \all~template built with a variable $\sigma_{\ell}$ within the range 0.02 to 0.62 \AA~($y$-axis). The CCFs tested at different $\sigma_{\ell}$ are showed in orange and the CCF corresponding to the default $\sigma_{\ell} \sim$0.43 used in this study (Table~\ref{tab:template}) is showed in red. The CCF corresponding to the same \dcep~spectrum and the G2 HARPS DRS template is showed as a dashed black curve and is shifted vertically for clarity. The four right plots show the behaviour of different CCF observables vs. $\sigma_{\ell}$ (from left to right and top to bottom; in red circles): {\it i}) the CCF $Q$-factor averaged over the 103 \dcep~spectra; {\it ii}) the same for the averaged \sncc~criterion; {\it iii}) the slope of the linear regression to the \dcep~BIS time series at a given $\sigma_{\ell}$ vs. the BIS time series at $\sigma_{\ell} \sim 0.43$; and {\it iv}) the same for the \rvc~time series. On the four plots, the red diamond corresponds to the default $\sigma_{\ell}$ used in this study for the \all~template and the black square corresponds to the G2 HARPS DRS template.}
\label{fig:line_width}
\end{figure*}

\subsection{Template line width}\label{res:line_width}

Here, we investigate the impact of the average line width of a given correlation template ($\sigma_{\ell}$) on the resulting CCFs and \vr.

\subsubsection{Impact of variable line width}

We first consider the impact of a variable $\sigma_{\ell}$ for one of our tailored correlation templates, \ie~the \all~template over the \gr~range. We cross-correlated the 103 HARPS-North spectra of \dcep~with the \all~template for different line widths ranging from $\sim$5\%~to 140\%~of the default \all~template line width, \ie~$\sigma_{\ell}$ ranging from $\sim$0.03 to $\sim$0.6 \AA~(Fig.~\ref{fig:line_width}). In terms of CCF shape, broadening the template lines has two results: first, it leads to a broadening of the CCF core (\ie~a shallower and wider CCF core); and second, it strongly reduces the noise or dispersion of the CCF wings. Both effects were already predicted by \ie~\cite{queloz95}. The decrease in CCF depth leads to a slight decrease of our CCF $Q$-factor averaged over the pulsation phase (yet still well over $Q = 4$). In contrast, the decrease of the CCF wing variability in the same time leads to a strong increase of our \sncc~criterion up to a maximum for $\sigma_{\ell}$ in the 0.35-0.45~\AA~range (Fig.~\ref{fig:line_width}, top right). This is the main reason for our choice of having wider lines in our tailored templates (Sect.~\ref{method:template}): broadening the template lines allows to make the CCF core more distinguishable from the CCF wings, even if the CCF core is made shallower. The decrease of the CCF wing variability with an increasing $\sigma_{\ell}$ also allows for a more robust CCF Gaussian or biGaussian modelling by making the CCF shape closer to that of a Gaussian \citep{queloz95}. We observe the same CCF behaviour as a function of $\sigma_{\ell}$ for our other templates (\deep, \medm, and \weak).

Increasing $\sigma_{\ell}$ (\ie~broadening the CCF) also reduces the CCF asymmetry variability and leads to smaller \vr~amplitudes (Fig.~\ref{fig:line_width}, middle and bottom right). We consider it a positive result as our focus is on increasing the consistency of Cepheid \vr: reducing the CCF asymmetry leads to less asymmetry-dependent \ie~less variable and more consistent \vr~time series (and thus less variable $p$-factors). On the contrary, Cepheid studies that focus on specific items such as the atmospheric velocity gradient should then use narrow-line templates to exacerbate the CCF asymmetry and enhance its impact on the \vr.

\subsubsection{Comparison with the G2 HARPS DRS default template}\label{res:line_width_G2}

We then compare our \all-template \dcep~CCFs with the CCFs computed from the same spectra but with the classical G2 template from the HARPS DRS. We retrieved the G2 template from the data made available by \cite{brahm17} and adapted it to our specifications (\ie~the \gr~wavelength domain and telluric exclusion ranges, Sect.~\ref{method:template}). Over our \gr~range, the G2 template includes more than 1700 very narrow lines ($\sigma_{\ell} \sim$0.08~\AA, Table~\ref{tab:template}). It gives significantly different CCFs with a narrower and deeper core and very flat wings (Fig.~\ref{fig:line_width}). The G2 CCF narrow core is induced by the small $\sigma_{\ell}$, while the CCF wing flatness is induced by the many blended lines and the occasional G2 template line mismatches \citep[][and see Fig.~\ref{fig:template}]{queloz95}. This leads to both a high CCF $Q$-factor and a high \sncc~criterion. Given that our templates are specifically tailored for Cepheids (\ie~much less line mismatches) and that our constraints on the line selection are much more stringent (no blended lines), our CCF wings are inevitably more noisy, justifying our use of wider template lines. On another hand, the G2 CCFs are much more asymmetric than the CCFs built from our \all~template and exhibit larger \vr~amplitudes (Fig.~\ref{fig:line_width}).

\subsection{Template line depth}\label{res:line_depth}

Here, we investigate the impact of the average line depth of the correlation template on the resulting Cepheid CCFs and \vr. We looked at our data obtained on our \gr~range only, with the four corresponding templates (\weak, \medm, \deep, and \all~templates), \ie~considering a sub-sample of 50 targets.

\begin{figure*}[ht!]
\centering
\includegraphics[width=1\hsize]{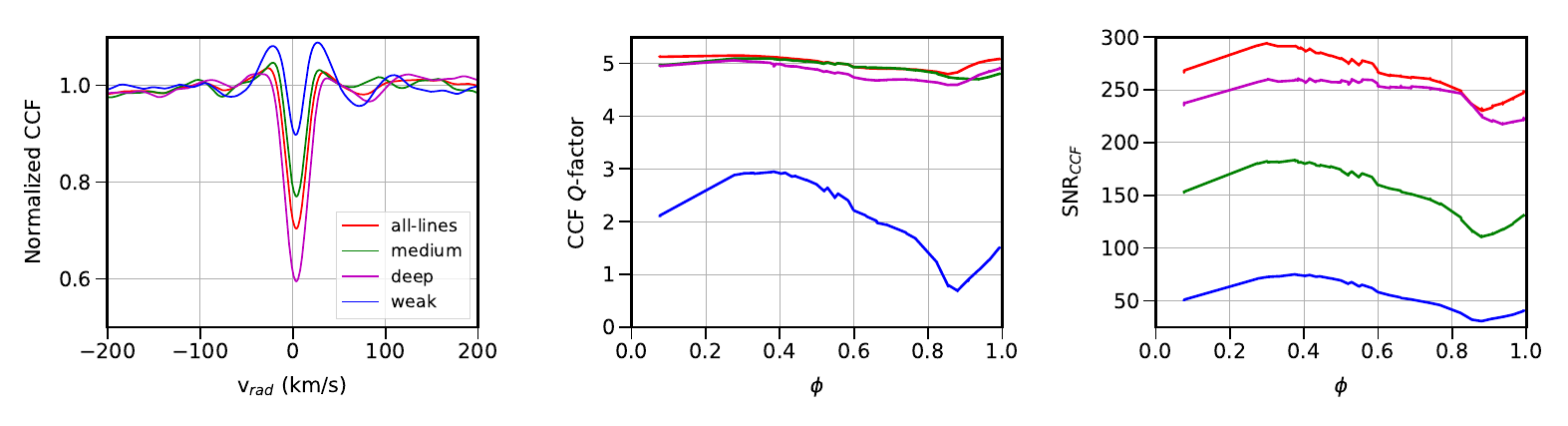}
\caption{HARPS \lcar~CCFs computed with our four depth-dependent correlation templates on the \gr~$\lambda$ range. Left: example of the four CCFs corresponding to a same spectrum and our four respective templates. Middle: CCF quality factor $Q$ of \lcar~vs. pulsation phase ($\phi$) for CCFs built based on our four templates (same colour code). Right: the same, for our CCF signal estimator \sncc.}
\label{fig:lcar_ccf}
\end{figure*}

\begin{figure*}[ht!]
\centering
\includegraphics[width=1\hsize]{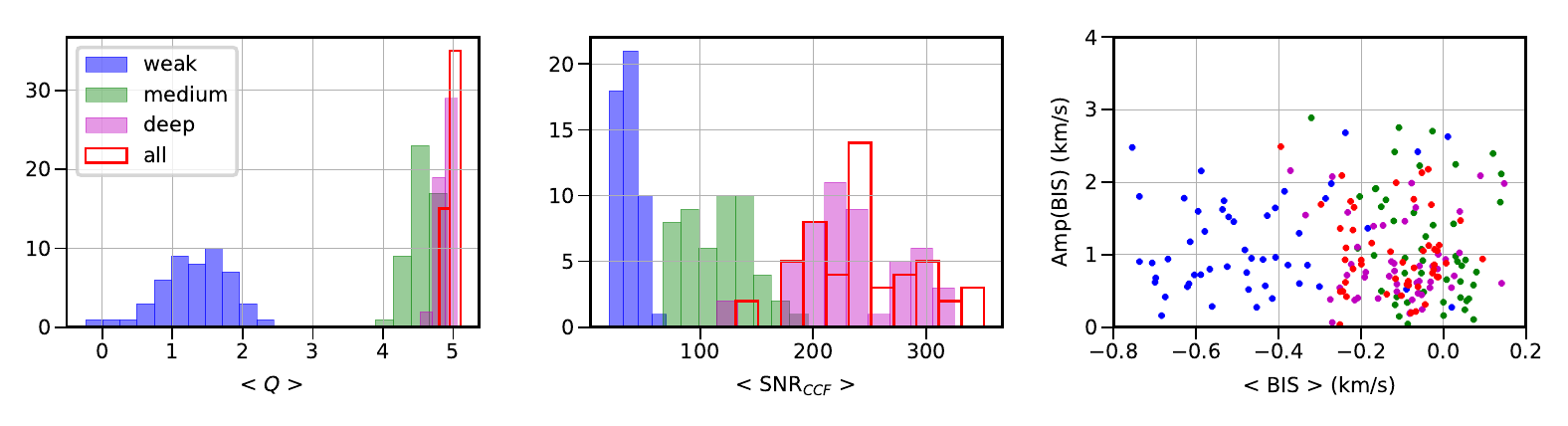}
\caption{CCF quality and asymmetry vs. correlation template. Left: histograms of the CCF $Q$ factor averaged over the pulsation phase for each target, for our four line depth-dependent templates. Middle: the same, for our CCF \sncc~proxy (same colour code). Right: amplitude of the BIS during the pulsation phase vs. averaged BIS for each target, for our four templates (same colour code).}
\label{fig:qual_template}
\end{figure*}

\subsubsection{CCF quality}

We look at the CCF quality for each of our four line depth-dependent correlation templates. As expected, the derived CCFs have a deeper core for templates corresponding to deeper lines (\ie~templates with a higher average line depth in Table~\ref{tab:template}), see Fig.~\ref{fig:lcar_ccf} (left). For three of our four templates, we generally find our CCF quality factor $Q$ to be nearly constant during the pulsation phase, with a value of $\sim$4-5 (Figs.~\ref{fig:lcar_ccf} and~\ref{fig:qual_template}). On the contrary, for the \weak~template, the $Q$ proxy is much smaller (between 0 and 3, Fig.~\ref{fig:qual_template} left) and significantly variable during the pulsation (Fig.~\ref{fig:lcar_ccf}, middle). We consider this as a criterion defining the good quality of a CCF and the reliability of the derived observables. As we already explained in Sect.~\ref{res:wave}, we empirically put a $Q = 4$ threshold to distinguish between high-quality ($Q \geq 4$) and mixed-quality ($Q < 4$) CCFs and data. Then, almost all our data derived from the \medm, \deep, and \all~templates meet our empirical quality threshold, while the data derived from the \weak~template are below this criterion. We note that even if using narrower lines for the \weak~template (as detailed in Sect.~\ref{res:line_width} for the \all~template), the derived $Q$-factors would still be below our $Q = 4$ empirical criterion. To obtain better-quality CCFs from such weak-line templates, we consider that the best solution would be to include more shallow lines, \ie~by alleviating our constraints on the line selection as described in Sect.~\ref{method:template}. Finally, we find our \sncc~parameter to be variable both during the pulsation phase (Fig.~\ref{fig:lcar_ccf}, right) and as a function of the correlation template used (Fig.~\ref{fig:qual_template}, middle). Understandably, the \sncc~values increase both for deeper lines and for templates including more lines.

\subsubsection{CCF asymmetry}

\begin{figure*}[ht!]
  \sidecaption
\includegraphics[width=13.5cm]{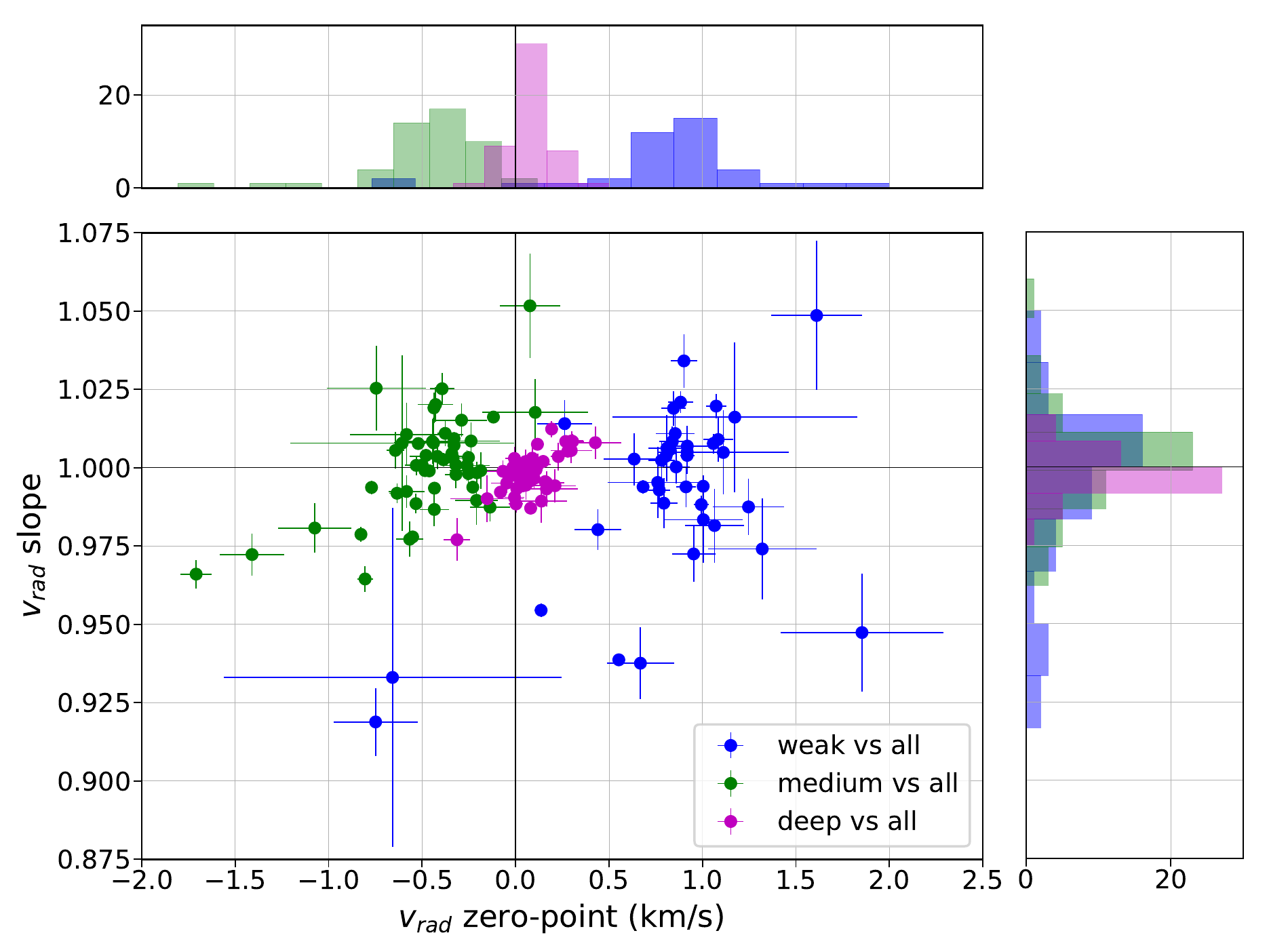}
\caption{Comparison of centroid \vr~computed with our respective correlation templates on the \gr~range. The figure represents the slope vs. zero-point distribution of the linear regression of the \weak~(blue), \medm~(green) and \deep~(purple)~\vr~vs. the \all~\vr, respectively.}
\label{fig:vrad_template}
\end{figure*}

We used our BIS variable to study the impact of the line depth on the CCF asymmetry. We do not find any clear pattern in terms of BIS amplitude over the pulsation phase $\phi$. However, we find the BIS averaged over $\phi$ to be higher (in absolute values) for the \weak~template compared to our three other templates (Fig.~\ref{fig:qual_template}, right). This would agree with shallower lines being more sensitive to the line asymmetry as reported by \cite{anderson16}.

\subsubsection{Radial velocities}

We finally look at the impact of the average line depth of the correlation template on the \vr. To do so, we compute for each of our selected targets three linear regressions: \weak~versus \all~\rvc, \medm~versus \all~\rvc, and \deep~versus \all~\rvc, respectively (on the \gr~wavelength range). We selected the centroid \vr~method as we derive it directly from the CCF and not from a CCF fit (as for Gaussian and biGaussian \vr~that are thus less robust). We display our results in Fig.~\ref{fig:vrad_template}.

We find that the zero-point of the linear regressions significantly change when going from one template to another. The median offset between \deep~and \all~\vr~time series is marginal ($0.07 \pm 0.12$ \kms), but the median offsets between \medm~and \all~\vr~and between \weak~and \all~\vr~are significant ($-0.43 \pm 0.31$~\kms~and $0.87 \pm 0.46$~\kms, respectively). Such \vr~offsets between correlation templates based on lines of different depth could be expected and agree with the findings of \eg~\cite{nardetto08,nardetto09,vasilyev17} on the dependency of Cepheid $\gamma$-velocities on the spectral lines and their depth. This would need a more in-depth analysis beyond the scope of this paper.

In contrast, the slopes of the linear regressions exhibit a significant variability but no clear trend from one template to another. The median slopes are comparable at a 1$\sigma$ level ($1.001 \pm 0.028$, $1 \pm 0.016$, and $0.998 \pm 0.006$ for the \weak~versus \all, \medm~versus \all, and \deep~versus \all~regressions, respectively). Yet, the so-called stellar velocity gradient (between spectral lines of different depths) is expected to have a significant impact on the Cepheid \vr~amplitudes, as detailed by \eg~\cite{nardetto06,nardetto07} for single lines. Significant differences in terms of CCF asymmetry and \vr~were also reported by \cite{anderson16} when cross-correlating spectra of \lcar~with two correlation templates (a weak-line one and a strong-line one, respectively, both having narrow lines). These authors reported an enhanced asymmetry for weak lines as well as an increased sensitivity or variability of their \vr. We consider that the present lack of significant trend (\ie~of the regression slope as a function of the average template line depth) is at least partially caused by our use of broader template lines (Sect.~\ref{res:line_width}), that reduce the asymmetry of the resulting CCFs and decrease the sensitivity of the \vr~time series. Yet, the slope distribution for the \weak~versus \all~regressions exhibits a much larger dispersion than for the \deep~versus \all~regressions. This would agree with strong lines showing less asymmetry and leading to more robust \vr.

\section{Conclusions}\label{sect:conclu}

We carried out a large spectroscopic survey of Classical MW Cepheids based on several thousands of high-resolution spectra from seven spectrographs. We detailed the framework that we implemented to derive and characterise Cepheid line profiles and \vr~time series as much consistent as possible based on the cross-correlation method. Briefly, the main steps of our formalism are the following:
\begin{enumerate}
\item normalising and standardising in wavelength all the spectra in the same way;
\item using pre-defined correlation templates with an emphasis on the selected lines, their average depth, and their width;
\item characterising not only the CCF Doppler shift, but its shape, depth, width, asymmetry and amount of extractable signal;
\item deriving the \vr~while accounting for the intrinsic CCF asymmetry.
\end{enumerate}
We show that each parameter and each step of the process has a significant impact on the derived Cepheid \vr: the wavelength range on which the spectrum is considered, the correlation template used for the cross-correlation (both in terms of line depth and template line width), and the way of computing the radial velocity have a significant impact on the derived \vr. For Baade-Wesselink studies, it basically means that significantly different projection factors and distances could be obtained for the same Cepheid, depending on how the former items are treated. Hence, giving at least a minimum of details on the \vr~computation process should be a pre-requisite for such studies. We also emphasise the importance of fully characterising the cross-correlation profile (CCF) through various estimators of its shape, width, depth, asymmetry and the amount of signal within it. We publish on-line both our tailored correlation templates, the derived CCFs, and the various \vr, line profile proxy, and CCF quality proxy time series computed from the CCFs. As we showed in this study, deriving fully consistent Cepheid \vr~from the cross-correlation method is not an easy task. Yet, it seems that the way towards more robust Cepheid \vr~(and thus more robust $p$-factors) goes through minimising the asymmetry of the line profile (here, the CCF) and reducing the sensitivity of the resulting \vr~to this asymmetry: \eg~using centroid \vr, favouring stronger lines in cross-correlation templates, or even using templates with somewhat broader lines than usual.

High-resolution spectroscopy of Cepheids has a lot of different applications. Our next objective is to take a fresh look at the computation of the Cepheid effective temperature \teff~from spectra. Cepheid \teff~exhibit a large variability (of the order of hundreds of Kelvin) over the pulsation phase, that has a significant impact on spectra. We aim at using our Cepheid sample to measure accurately both the absolute average Cepheid \teff~and its variation over the pulsation cycle by rehashing pre-existing methods based on the ratio of the depths or EWs of selected line pairs \citep[see \eg][respectively]{kovtyukh00,sousa10}. Other stellar parameters can be easily derived from Cepheid spectra, such as \eg~metallic abundances \citep{luck18}.

Another possibility is to use again the CCF as the spectrum's proxy to deduce other parameters than the \vr. An interesting point is to build a model grid of Cepheid CCF, in the same way as done by \cite{britavskiy17} with their grid of synthetic bisectors. In the case of non-pulsating stars, CCF profiles can already be used (instead of the spectra themselves) to derive \eg~the \teff, \logg~and metallicities \citep{malavolta17}.

Finally, it is interesting to see if and how the Cepheid CCF formalism that we detailed here can be applied to the data of the {\it Gaia Radial Velocity Spectrograph} \citep[RVS,][]{cropper18}. The RVS spectra are centred on the Calcium II near-infrared triplet, on a narrow wavelength band if compared to the spectrographs implemented within this study. Yet, they are expected to produce transit \vr~measurements with a precision well below 1 \kms, \ie~enough to scan the Cepheid \vr~pulsation amplitude \citep{sartoretti18,katz18}. This would thus be a great opportunity given that {\it Gaia} will observe several thousands of MW Cepheids.

\begin{acknowledgements}
  The authors acknowledge the support of the French Agence Nationale de la Recherche (ANR) under grant ANR-15-CE31-0012-01 (project UnlockCepheids). WG and GP gratefully acknowledge support from the Chilean Centro de Astrofisica y Tecnologias Afines (CATA) BASAL grant AFB-170002. WG also acknowledges support from the Chilean Ministry of Economy, Development and Tourism's Millennium Science Iniciative through grant IC120009 awarded to the Millenium Institute of Astrophysics (MAS). The research leading to these results has received funding from the European Research Council (ERC) under the European Union's Horizon 2020 research and innovation program (grant agreement No. 695099). The Swiss 1.2m Euler telescope and the CORALIE spectrograph are supported by the Swiss National Science Foundation. This research is partly based on observations made with the Mercator Telescope, operated on the island of La Palma by the Flemish Community, at the Spanish Observatorio del Roque de los Muchachos of the Instituto de Astrof\'isica de Canarias. HERMES is supported by the Fund for Scientific Research of Flanders (FWO), Belgium, the Research Council of K.U. Leuven, Belgium, the Fonds National de la Recherche Scientifique (F.R.S.-FNRS), Belgium, the Royal Observatory of Belgium, the Observatoire de Gen\`eve, Switzerland, and the Th\"uringer Landessternwarte, Tautenburg, Germany. 
This research made use of the SIMBAD and VIZIER databases at the CDS, Strasbourg (France), and NASA's Astrophysics Data System Bibliographic Services. We made use of the \texttt{Python} programming language \citep{rossum95} and the open source \texttt{Python} packages \texttt{numpy} \citep{vanderwalt11}, \texttt{scipy} \citep{jones01}, \texttt{matplotlib} \citep{hunter07}, and \texttt{astropy} \citep{astropy13}. The authors would like to thank the anonymous referee for the useful remarks and points, especially on using the cross-correlation method.
\end{acknowledgements}

\bibliographystyle{aa}
\bibliography{hrspips}

\begin{appendix}
\section{Detail of our Cepheid sample}\label{appdix:sample}

\renewcommand{\arraystretch}{1.25}
\begin{center}
\onecolumn
\setlength\tabcolsep{3.5pt}
\begin{longtable}[1]{l l c c | c c | c c c r}\\
\caption{\label{tab:sample} Detail of our Cepheid sample. Columns 2 to 4 give the pulsation period $P$, the yearly period drift $dP/dT$ (if known) and the adopted reference epoch for period phasing T$_{0}$. Column~5 indicates if the target is a clear (single-lined) spectroscopic binary based on its \vr~($y$) or if there is some \vr~scatter ($s$) that make it a possible SB1 (based on the present data only; other Cepheids in the list are known binaries but with no significant effect on our present \vr~time series). In case of a SB1, column~6 gives the most recent reference providing the best companion orbital parameters, if any. Columns 7 to 9 give the total number of spectra $N_{\rm sp}$ used in this study (all spectrographs included), the corresponding number of distinct observation epochs $N_{\rm epoch}$ and the deduced pulsation phase coverage ($\phi$ cov.). Column 10 give the detail of the spectrum distribution per instrument for each target.}\\
\hline
Cepheid & $P$ & $\frac{dP}{dT}$ & T$_{0}$ & SB1 & Ref.$^{(1)}$ & $N_{\rm sp}$ & $N_{\rm epoch}$ & $\phi$ cov. & Instr.$^{(2)}$ \\
        & [day] & [s.yr$^{-1}$] & [MJD]   &     &      &              &                 & [\%]    & \\
\hline
\hline
\endfirsthead
\caption{Continued.}\\
\hline
Cepheid & $P$ & $\frac{dP}{dT}$ & T$_{0}$ & SB & Ref. & $N_{\rm sp}$ & $N_{\rm epoch}$ & $\phi$ cov. & Instr. \\
        & [day] & [s.yr$^{-1}$] & [MJD]   &    &      &              &                 & [\%]        &
          \\
\hline
\hline
\endhead
\hline
\endfoot
SU Cas & 1.949 & -0.02 & 48347.23 & y      &      & 12  & 12 & 27 & S (12). \\
BP Cir & 2.398 &       & 54498.29 & s      & -    & 34  & 20 & 60 & C (27), H (6), F (1). \\
AV Cir & 3.065 &       & 53617.3  & s      & -    & 161 & 21 & 33 & Ur (100), Ub (60), F (1). \\
V1334 Cyg & 3.333 &    & 40124.033& y      & G18b & 70  & 49 & 93 & HM (56), S (14). \\
BG Cru & 3.343 &       & 40393.16 & s      & -    & 159 & 32 & 54 & Ur (72), Ub (59), C (20), F (4), H (4). \\
R Tra  & 3.389 &       & 40837.71 &        &      & 15  & 11 & 33 & H (14), F (1). \\
Y Car  & 3.64  &       & 41040.89 & y      & -    & 43  & 24 & 54 & C (35); H (6), F (2). \\
RT Aur & 3.728 & -0.08 & 47956.89 &        &      & 138 & 17 & 27 & Ur (81); Ub (54), S (3). \\
SU Cyg & 3.846 &       & 54322.36 & y      & This study & 13  & 12 & 33 & S (13). \\
AH Vel & 4.227 &       & 42035.175& s      & -    & 151 & 36 & 54 & Ur (72), Ub (48), C (19), H (8), F (4). \\
Y Lac  & 4.324 &       & 41746.245&        &      & 14  & 12 & 33 & S (14). \\
T Vul  & 4.435 &       & 52242.743&        &      & 18  & 16 & 36 & S (17), H (1). \\
FF Aql & 4.471 & 0.12  & 47405.98 & y      & G18a & 95  & 88 & 93 & C (44), HM (36), S (13), H (2). \\
S Cru  & 4.69  &       & 34973.02 &        &      & 13  & 12 & 30 & H (12), F (1). \\
VZ Cyg & 4.864 &       & 41705.202&        &      & 18  & 17 & 51 & S (18). \\
V350 Sgr& 5.154&       & 35316.727& y      & G18a & 27  & 19 & 42 & C (21), H (6). \\
AX Cir & 5.273 &       & 38199.04 & s      & -    & 137 & 41 & 66 & Ur (64), Ub (40), C (22), H (9), F (2). \\
$\delta$ Cep & 5.366 & -0.0765 & 48304.772 &  &   & 106 & 40 & 87 & HN (103), S (3). \\
X Lac  & 5.445 &       & 42737.632&        &      & 11  & 11 & 33 & S (11). \\
V659 Cen& 5.623&       & 53483.1  & y      & -    & 20  & 17 & 42 & C (16), H (3), F (1). \\
MY Pup & 5.695 &       & 41043.22 & s      & -    & 123 & 17 & 30 & Ur (72), Ub (48), F (3). \\
Y Sgr  & 5.77  & 0.14  & 48700.297& y      & -    & 36  & 27 & 57 & H (22), C (14). \\
EW Sct & 5.824 &       & 49705.227& y      & -    & 73  & 23 & 36 & Ur (40), Ub (30), F (3). \\
FM Aql & 6.114 &       & 35151.223&        &      & 12  & 11 & 33 & S (10), H (2). \\
AW Per & 6.464 &       & 42708.559&        &      & 9   & 9  & 24 & S (9). \\
U Sgr  & 6.745 & 0.11  & 48336.5  &        &      & 86  & 26 & 36 & Ur (48), Ub (36), H (2). \\
V636 Sco & 6.797 &     & 40363.892& y      & G18a & 98  & 40 & 54 & Ur (40), Ub (30), C (21), H (6), F (1). \\
V496 Aql & 6.807 &     & 56715.564& y      & This study & 34  & 32 & 69 & C (14), S (14), H (6). \\
X Sgr  & 7.013 & 0.1665& 48707.742& s      & -    & 49  & 31 & 66 & H (30), C (17), F (2). \\
U Aql  & 7.024 & -0.12 & 51000.122& y      & G18a & 56  & 42 & 69 & C (30), S (18), H (8). \\
$\eta$ Aql & 7.177 & -0.1064 & 48069.417 &  &     & 13  & 12 & 39 & S (13). \\
R Mus  & 7.51  &       & 26495.788& s      & -    & 104 & 32 & 54 & Ur (56), Ub (32), C (12), H (3), F (1). \\
V440 Per& 7.57 &       & 44869.44 &        &      & 8   & 7  & 24 & S (8). \\
W Sgr  & 7.595 & 0.0573& 48690.7  & y      & G18a & 89  & 81 & 90 & C (67), HM (18), H (4). \\
U Vul  & 7.991 & 0.61  & 48311.157& y      & -    & 32  & 32 & 60 & S (17), C (13), H (2). \\
DL Cas & 8.0   & 0.31  & 48300.049& y      & -    & 16  & 15 & 30 & S (16). \\
V636 Cas & 8.377 &     & 44000.855&        &      & 8   & 8  & 21 & S (8). \\
S Sge  & 8.382 & 0.09  & 48311.094& y      & -    & 32  & 27 & 54 & C (17), S (15). \\
S Mus  & 9.66  &       & 40298.92 & y      & G18a & 117 & 38 & 66 & Ur (56), Ub (32), C (19), H (8), F (2). \\
S Nor  & 9.754 & 0.47  & 48320.068&        &      & 105 & 33 & 51 & Ur (49), Ub (28), C (21), H (6), F (1). \\
$\beta$ Dor & 9.843 & 0.4305 & 50274.926 &  &     & 185 & 27 & 48 & Ub (78), Ur (60), H (46), F (1). \\
$\zeta$ Gem & 10.15 & -1.5362 & 48708.069 & &     & 189 & 31 & 48 & Ub (80), Ur (60), H (47), S (2). \\
RX Aur & 11.623 &      & 39075.13 &        &      & 9   & 9  & 27 & S (9). \\
TT Aql & 13.755 & 0.64 & 48308.557&        &      & 72  & 22 & 33 & Ur (40), Ub (30), H (2). \\
TX Cyg & 14.71  &      & 43794.471&        &      & 7   & 7  & 21 & S (7). \\
UZ Sct & 14.749 &      & 41101.0  &        &      & 20  & 17 & 48 & F (20). \\
AV Sgr & 15.415 &      & 44016.75 & s      & -    & 22  & 19 & 45 & F (22). \\
X Cyg  & 16.386 & 1.09 & 48319.576&        &      & 17  & 14 & 39 & S (17). \\
RW Cam & 16.414 &      & 37389.07 &        &      & 7   & 7  & 21 & S (7). \\
CD Cyg & 17.074 & 4.59 & 48321.236&        &      & 9   & 9  & 27 & S (9). \\
Y Oph  & 17.126 & -2.9972 & 46224.13 & s   & -    & 26  & 26 & 60 & C (11), S (8), H (7). \\
YZ Car & 18.168 &      & 53091.13 & y      & G18a & 17  & 14 & 48 & C (10), H (6), F (1). \\
RU Sct & 19.704 & 5.58 & 48335.591&        &      & 58  & 18 & 33 & Ur (34), Ub (22), F (1), H (1). \\
RZ Vel & 20.398 &      & 34845.07 &        &      & 78  & 25 & 48 & Ur (42), Ub (23), H (12), F (1). \\
V340 Ara & 20.814 &    & 52351.25 &        &      & 15  & 14 & 42 & F (15). \\
WZ Car & 23.018 &      & 44142.67 &        &      & 80  & 20 & 30 & Ur (50), Ub (30). \\
VZ Pup & 23.175 &      & 41120.69 &        &      & 89  & 23 & 27 & Ur (55), Ub (34). \\
T Mon  & 27.025 & 4.96 & 43783.791 &       &      & 85  & 22 & 39 & Ur (49), Ub (28), S (7), H (1). \\
$\ell$ Car & 35.558 & 33.4072 & 50583.743 &  &    & 253 & 43 & 72 & H (111), Ub (79), Ur (62), F (1). \\
U Car  & 38.829 & 86.45& 48336.858 &       &      & 108 & 36 & 54 & Ur (56), Ub (32), C (13), H (6), F (1). \\
RS Pup & 41.461 & 152.0691 & 48323.909 &   &      & 82  & 27 & 51 & Ur (42), Ub (24), H (16). \\
SV Vul & 44.993 & -263.98 & 48307.729  & s & -    & 33  & 31 & 69 & S (31), H (2). \\
V1496 Aql & 65.77 &    & 53911.0  &        &      & 74  & 18 & 24 & Ur (50), Ub (24). \\
S Vul  & 68.54  & -967.4 & 48333.28 &      &      & 29  & 27 & 63 & S (27), H (2). \\
\hline
\end{longtable}
$(1)$ G18a, G18b stand for \cite{gallenne18b} and \cite{gallenne18a}. $(2)$ H, S, C, F, HM, and HN stand for HARPS, SOPHIE, CORALIE, FEROS, HERMES, and HARPS-North, respectively. Ub and Ur stand for the blue and red arms of UVES.
\twocolumn
\end{center}
\renewcommand{\arraystretch}{1}

\section{Details on the computation of CCF, radial velocities and observables}\label{appdix:method}

\subsection{CCF computation}

We use the \texttt{crosscorrRV} package from the Python AstroLib library \footnote{\url{http://www.hs.uni-hamburg.de/DE/Ins/Per/Czesla/PyA/PyA/pyaslDoc/pyasl.html}} to cross-correlate each observed spectrum with a given correlation template. We compute the CCF on a default \vr~grid ranging from $-200$ to $+200$~\kms~with a 1~\kms~step. Such a step is in the order of what is commonly done in the DRS of the spectrographs considered in this study \citep[see \eg~][]{queloz95,baranne96}, and it is enough to provide precise \vr~measurements (\eg~through a CCF Gaussian or biGaussian fit). The \texttt{crosscorrRV} package is an easy-to-use and time-tested framework for cross-correlating 1D spectra with templates. The correlation template is successively Doppler-shifted in wavelength, and then linearly interpolated at the spectrum wavelength points for the correlation. Given that our correlation templates are sampled with a $\sim$0.02 \AA~step (roughly corresponding to a $\sim$1~\kms~step in terms of \vr~and similar to the wavelength step of our input spectra), we do not expect this linear interpolation to modify our template lines and to impact the derived observables.

\subsection{Observable computation}

\paragraph{CCF continuum and depth --} We denote the \vr~extension of the CCF core (\ie~the area of the CCF peak located below the CCF lower shoulder, see Fig.~\ref{fig:param_ccf}) as \dcore~and we denote the reunion of the CCF wing \vr~ranges (\ie~the \vr~ranges both left and right of the two CCF shoulders) as \dwing. We thus define the CCF continuum \coo~as
\[
\mathcal{C}^{\rm o} = \ < \mathrm{CCF}  [\Delta_{\rm wings}] > 
\]
Next, we define the CCF contrast (normalised here by the CCF continuum) as
\[
CT = 1 - \frac{\mathrm{min} \left( \mathrm{CCF} [\Delta_{\rm core}] \right) }{\mathcal{C}^{\rm o}}
\]
while we define the (normalised) CCF core depth ($D$) as
\[
D = \frac{\mathrm{max} \left( \mathrm{CCF} [\Delta_{\rm core}] \right) - \mathrm{min} \left( \mathrm{CCF} [\Delta_{\rm core}] \right)}{\mathcal{C}^{\rm o}}
\]
\paragraph{CCF core equivalent width (EW) --} We compute EW through a direct integration of the CCF over \dcore:
\[
\mathrm{EW} = \int_{\Delta_{\rm core}} \! \left(1 - \frac{\mathrm{CCF}[v_{\rm rad}]}{\mathrm{max} (\mathrm{CCF}[\Delta_{\rm core}])}\right) \, \mathrm{d}(v_{\rm rad})
\]
The integration is done on a finer \vr~grid (\vr~step of 50 \ms) than the CCF computation (see above). The CCF EW is equivalent to the width (in \kms) of a theoretical rectangle with a height equal to 1 (considering the normalised CCF) and with a surface equal to the area covered by the CCF core.

\paragraph{CCF first moment --} To derive the first moment of the CCF core \rvc, we compute the cumulated integral of the CCF profile over \dcore, using the same \vr~grid as for the CCF EW. Integrating the CCF core including the small area above the lower shoulder and below the higher shoulder does not lead to significant changes to the CCF EW and \rvc. The EW values are marginally higher (by $\sim$1-2\%~or typically 100 \ms, and the \rvc~amplitudes increase only marginally, by a few tens of \ms~(\ie~by a value lower than the typical \rvc~uncertainties). We adopt the same nomenclature as of \cite{nardetto06,nardetto09} for the different \vr~computation methods applied to the CCF: \rvc~for the CCF first moment (by analogy with $RV_{\rm c}$ for single-line first moments), \rvg~for the CCF Gaussian (by analogy with $RV_{\rm g}$ single-line Gaussian models), etc.

\paragraph{CCF bisector --} We compute the CCF bisector by dividing the CCF core into 100 horizontal slices (between the minimum and the maximum of the CCF peak) and by computing the mean \vr~for each slice. We then compute the corresponding Bisector Inverse Span (BIS), defined as the \vr~span between a top and a bottom domain of the bisector \citep{queloz01b}. If denoting $V_{\rm top}$ and $V_{\rm btm}$ as the mean \vr~of these top and bottom BIS domains, we have BIS~$=$~$V_{\rm top} - V_{\rm btm}$. We use the top and bottom bisector region definition given by \cite{galland05}: \ie~a top region extending from 15 to 46\%~of the CCF depth $D$ and a bottom region from 57 to 85\%~of $D$.

\paragraph{CCF quality proxies --} We compute our CCF quality factor $Q$ as 
\[
Q = \frac{CT}{\sigma \left( \mathrm{CCF} - \mathrm{CCF_{smth}} \right)}
\]
where $CT$ represents the CCF contrast, $\sigma$ denotes the standard deviation, and CCF$_{smth}$ represents the CCF smoothed over a \vr~range equal to 2.355~$\times$~\dcore. Then we compute our CCF S/N estimator as
\[
\mathrm{SNR_{\rm CCF}} = \frac{D}{\sigma \left( \mathrm{CCF} [\Delta_{\rm wings}] \right)}.
\]
As shown in Sect.~\ref{sect:results}, $Q$ does not necessarily depend linearly on the CCF contrast $CT$. It stays on a plateau at $\sim$4-5 for all good-quality CCFs but starts to significantly decrease when the CCF wings become noisy and the CCF core very shallow.

\paragraph{CCF Gaussian fit --}
We model the profile of the CCF core or main peak (considered over its \vr~extension \dcore) with a four-parameter Gaussian function ($\mathcal{G}$):
\[
\mathcal{G} \ (v_{\rm rad}) = \mathrm{C}_{\rm g} \times \, \left(1 - D_{\rm g} \ \exp{(-4 \ln{2} \frac{(v_{\rm rad} - RV_{\rm cc-g})^{2}}{\mathcal{F}^{2}})}\right)
\]
where \coo$_{\rm g}$ refers to the offset of the Gaussian function, $D_{\rm g}$ to its (normalised) depth, $\mathcal{F}$ to its FWHM, and \rvg~to the Gaussian \vr. 

\begin{figure*}[ht!]
\centering
\includegraphics[width=0.49\hsize]{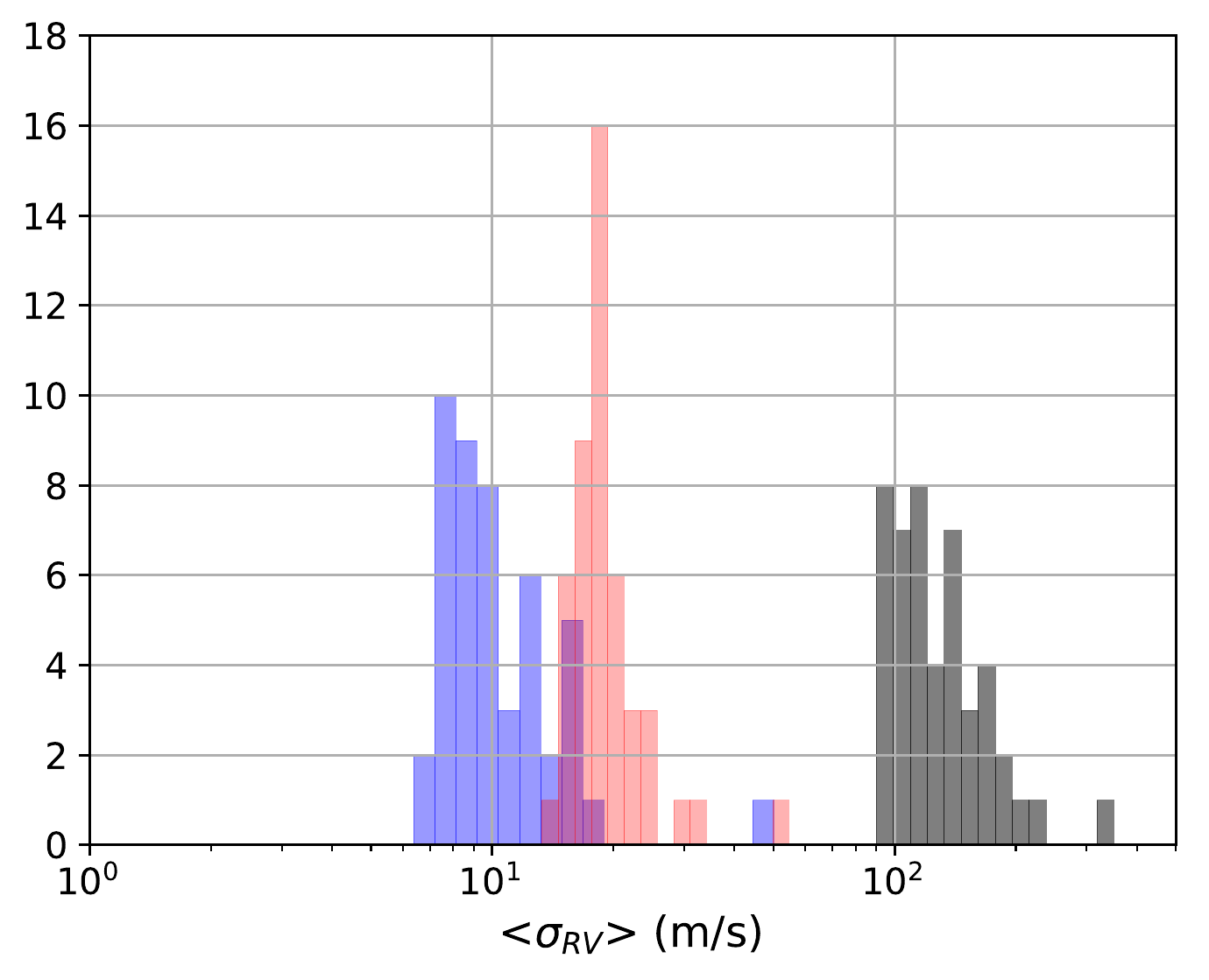}
\includegraphics[width=0.49\hsize]{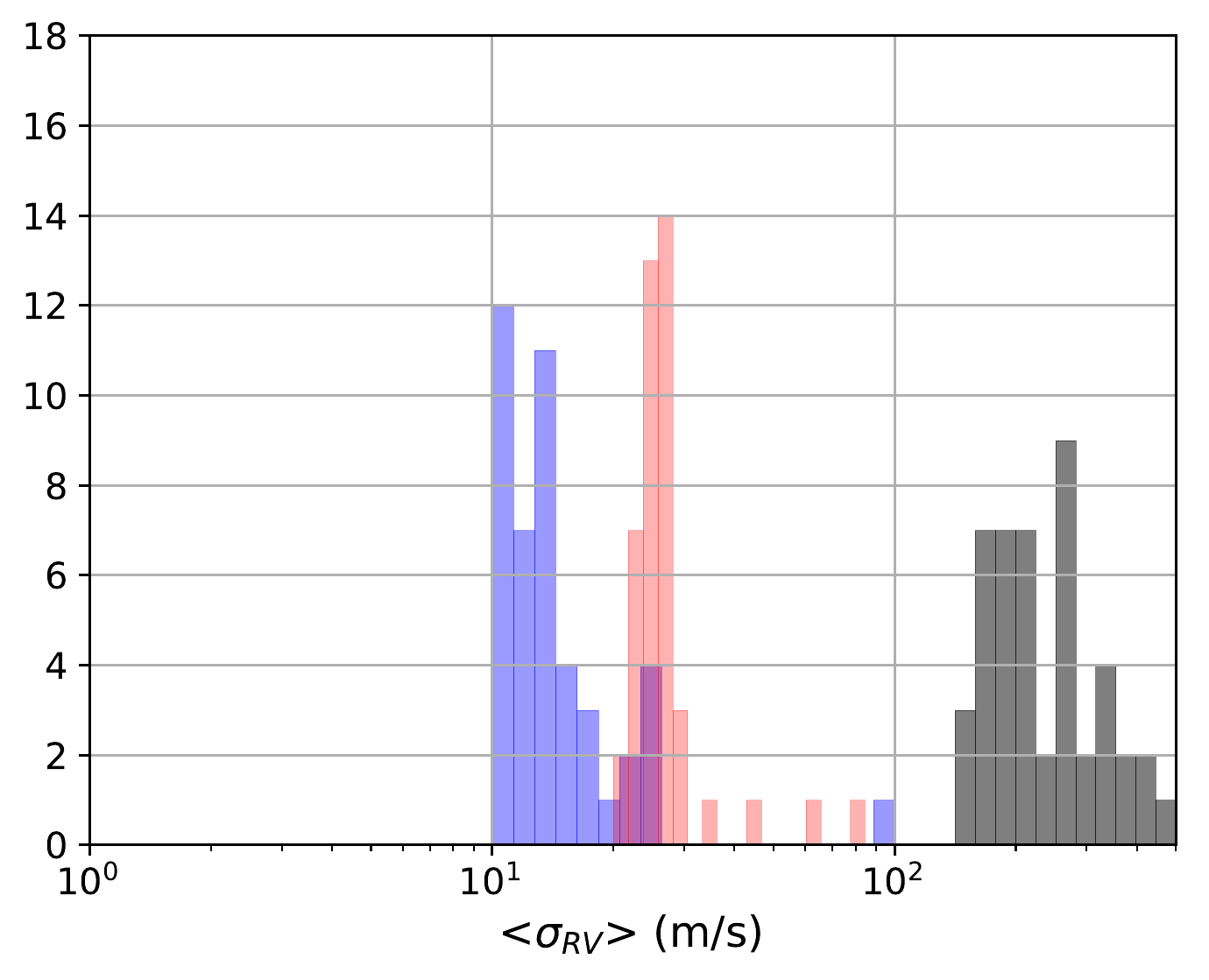}
\caption{Distribution of the mean \vr~uncertainties (\ie~averaged over all the \vr~measurements for each target), for the \vr~computed on the \gr~$\lambda$ range. Left: \all~template; right: \medm~template. The uncertainty distributions are displayed in blue, red and black shades for \rvg, \rvbg~and \rvc, respectively.}
\label{fig:unc_rv}
\end{figure*}

\begin{figure*}[ht!]
\centering
\includegraphics[width=1.\hsize]{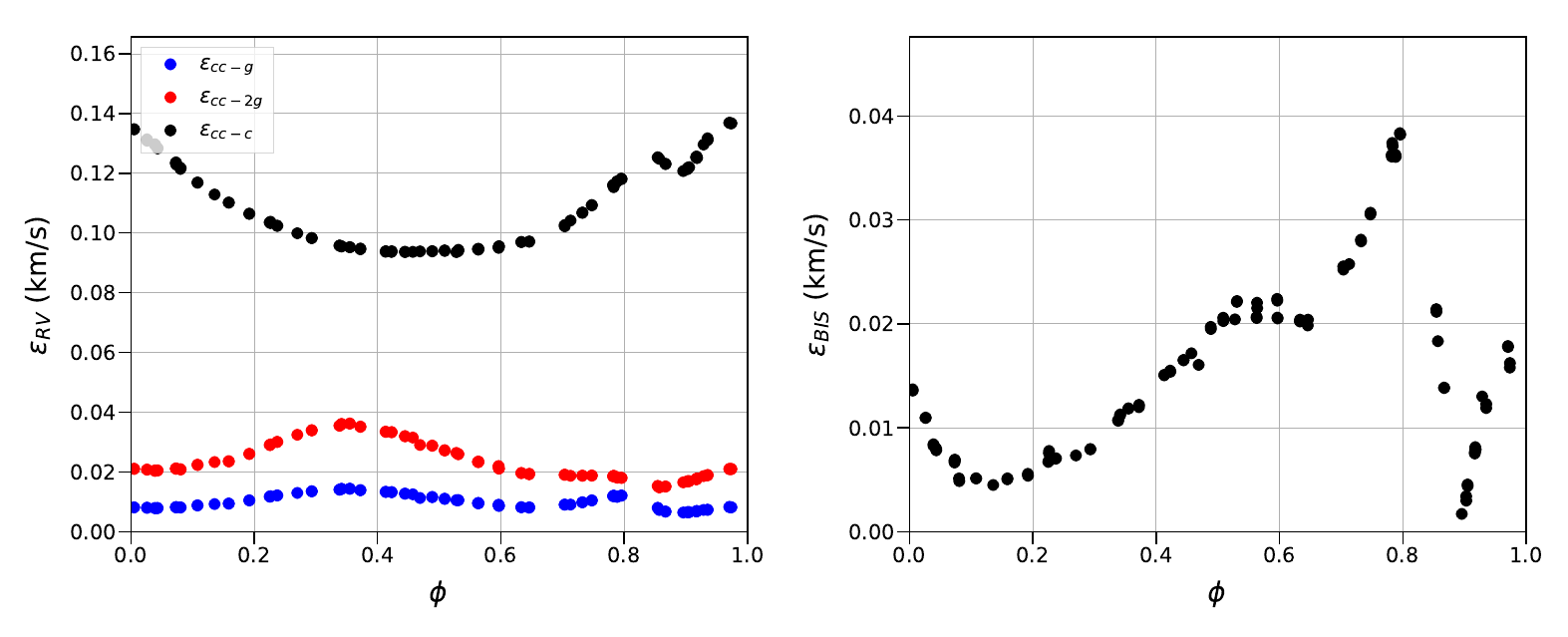}
\caption{Uncertainties vs. pulsation phase. Based on HARPS-North \dcep~spectra cross-correlated with the \all~template on the \gr~$\lambda$~range. Left: \vr~uncertainties vs. pulsation phase ($\phi$), for Gaussian, biGaussian and centroid \vr~(blue, red and black dots, respectively); right: BIS uncertainties vs. phase.}
\label{fig:unc_phi}
\end{figure*}

\paragraph{CCF biGaussian fit --}
As done with the Gaussian model, we fit the CCF core with a five-parameter biGaussian function ($\mathcal{B}$):
\[
\mathcal{B} \ (v_{\rm rad}) = \mathrm{C}_{\rm 2g} \times \, \left(1 - D_{\rm 2g} \ \exp{(-4 \ln{2} \frac{(v_{\rm rad} - RV_{\rm cc-2g})^{2}}{\mathcal{F}_{\rm i}^{2}})}\right)
\]
where $\mathcal{F}_{\rm i} = \mathcal{F}_{\rm L}$ if \vr~$<$~\rvbg~(equivalent to the FWHM of a Gaussian model of the CCF blue wing) and $\mathcal{F}_{\rm i} = \mathcal{F}_{\rm R}$ if \vr~$>$~\rvbg~(FWHM of a Gaussian fitted to the CCF red wing). Other parameters (continuum, depth, \vr) are analogous to the Gaussian ones. We point out that our biGaussian model is slightly different from the one used by \cite{nardetto06}, as the radial velocity \rvbg~is one of the free parameters. For both the Gaussian and biGaussian fits, we use a non-linear Least-Square method as implemented in the \texttt{curve$\_$fit} function of the \texttt{scipy.optimize} Python package.

\subsection{Uncertainties}

\begin{figure*}[ht!]
  \centering
  \includegraphics[width=1.\hsize]{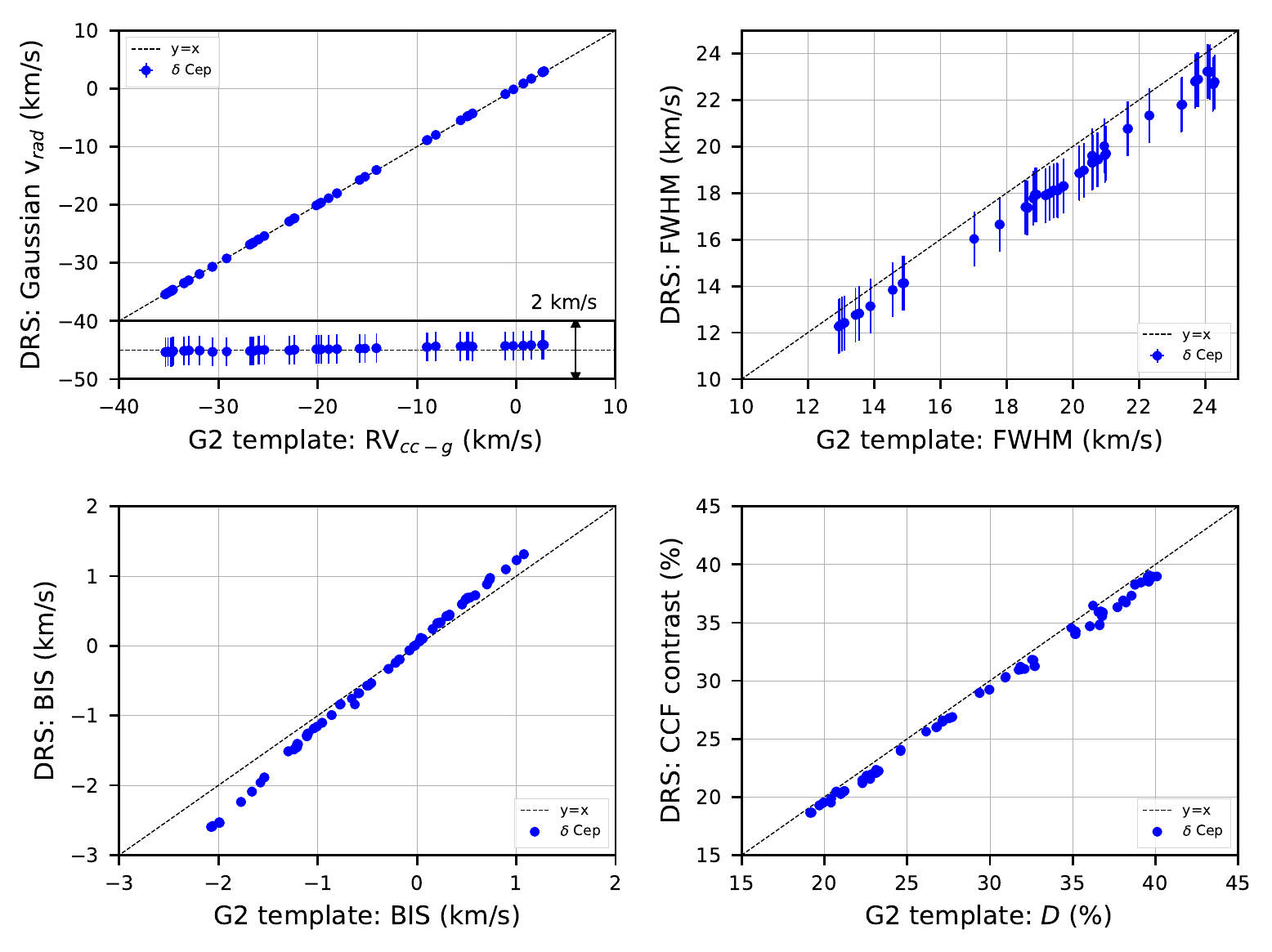}
\caption{Comparison between our \texttt{s1d}-based CCF computation and the automatic (\texttt{e2ds}-based) DRS CCF computation (see text). For each plot, the data that we obtain from the cross-correlation of the \dcep~HARPS-North (1D) spectra with the G2 template adapted to our \gr~wavelength range (see Sect.~\ref{res:line_width_G2}) are plotted against the corresponding observables retrieved from the available DRS data (automatically produced from the 2D spectra and the default G2 template). From top to bottom and left to right: Gaussian \vr, Gaussian FWHM, CCF BIS, and CCF depth.}
\label{fig:comp_G2}
\end{figure*}

While choosing which \vr~computation method(s) to use for Cepheids is an important question, computing reliable and realistic uncertainties on the CCF \vr~and other line profile observables is also important. Typically, the DRS of high-resolution spectrographs include two main sources of uncertainty on the \vr: (1) the photon noise (proportional to the inverse of the spectrum's S/N); and (2) the instrumental noise (or read-out noise), corresponding to the instrumental \vr~precision and the instrumental stability \citep{baranne96,pepe02}. For the best spectrographs (\eg~HARPS), the increased instrumental stability (Table~\ref{tab:list_spectro}) and easily reachable high spectrum S/N mean that \vr~uncertainties lower than 1~\ms~on average can be routinely achieved \citep{pepe18}. Such \vr~precision makes sense \eg~for exoplanet surveys around (generally) non-pulsating stars, for which the CCF can be properly fitted with a Gaussian model. However, Cepheids (and other radially pulsating stars) pose a different kind of challenge, due to the Cepheid \vr~accuracy being strongly dependent on the correlation template and the \vr~computation method used. 

We decided here to compute our \vr~(and other line profile observable) uncertainties based on our CCFs and their characteristics to remain consistent within our CCF-based framework, instead of using the spectrum's photon noise. We use \sncc~(see above) to estimate the uncertainty on the centroid radial velocity \rvc:
\[
\epsilon_{\rm cc-c} = \frac{W}{\rm SNR_{\rm CCF}} \, = \frac{W}{D} \times \sigma \left( \mathrm{CCF} [\Delta_{\rm wings}] \right),
\]
where $W$ denotes the width of the CCF core at half its depth $D$. Such a formula is analogous in terms of dimensions to the formulae provided by \eg~\cite{queloz95,baranne96}. It makes the \rvc~uncertainty directly dependent on the correlation template used to compute the CCF and on the observational pulsation phase. 
Next, to derive the BIS uncertainty ($\epsilon_{\rm BIS}$), we first compute the uncertainties on $V_{\rm top}$ and $V_{\rm btm}$ as classical errors on the mean (\ie~as the ratio of the bisector \vr~dispersion in the defined top and bottom regions over the square root of the number of \vr~points in the two respective bisector slices). Then we add quadratically these two uncertainties to obtain $\epsilon_{\rm BIS}$.

For all the observables (\vr, Gaussian FWHM, biGaussian asymmetry, etc) derived from a Gaussian or a biGaussian CCF model, we use the uncertainties derived within the fit, \ie~the 1$\sigma$~uncertainties computed from the square root of the covariance matrix diagonal. These uncertainties describe how closely the Gaussian or biGaussian fit agrees with the modeled CCF. We thus consider it valid to use them within our CCF-based formalism. We compare the derived uncertainties on \rvc~($\epsilon_{\rm cc-c}$, from our formula), \rvg~and \rvbg~($\epsilon_{\rm cc-g}$ and $\epsilon_{\rm cc-2g}$ from the fits) in Fig.~\ref{fig:unc_rv}. Overall, $< \epsilon_{\rm cc-2g} >$ is twice larger than~$< \epsilon_{\rm cc-g} >$ ($\sim$20 \ms~against $\sim$10 \ms, respectively, for the \all~correlation template), which agrees with the biGaussian model being more sensitive to the asymmetry and somewhat less robust than the Gaussian model. Meanwhile, $< \epsilon_{\rm cc-c} >$ is about one order of magnitude larger than~$< \epsilon_{\rm cc-g} >$. We cannot infer anything from this given that our computation formula for $< \epsilon_{\rm cc-c} >$ is arbitrary (see above), but we note that the shape of the distribution is similar for $< \epsilon_{\rm cc-c} >$ and $< \epsilon_{\rm cc-g} >$, which gives us confidence in our formula. Finally, going from a template with more and deeper lines to a template with less and shallower lines leads understandably to an increase of the \vr~uncertainties (Fig.~\ref{fig:unc_rv}).

We also display in Fig.~\ref{fig:unc_phi} the \vr~and BIS uncertainties versus the pulsation phase for one of our targets (\dcep). The Gaussian and biGaussian \vr~uncertainties are slightly variable with the phase, while the centroid \vr~uncertainties ($\epsilon_{\rm cc-c}$) are significantly variable (being an order of magnitude larger than the other uncertainties). The $\epsilon_{\rm cc-c}$ uncertainties do not exhibit the same pattern of variation as $\epsilon_{\rm cc-g}$ and $\epsilon_{\rm cc-2g}$. The BIS uncertainties are significantly variable with $\phi$, with $\epsilon_{\rm BIS}$ being the highest for the largest CCF asymmetry (see Fig.~\ref{fig:catalogue}). In average, $\epsilon_{\rm BIS}$ has the same order of magnitude as $\epsilon_{\rm cc-g}$ or $\epsilon_{\rm cc-2g}$. We note that our $\epsilon_{\rm BIS}$ estimation may be a conservative one as the BIS is a \vr~differential measurement \citep{anderson19}. 

\subsection{Comparison with the DRS-based CCF computation}

Here, we compare our \texttt{s1d}-based CCF computation and derived observables with the \texttt{e2ds}-based CCF computation performed automatically by the spectrograph DRS. We consider on the one hand the 103 HARPS-North \dcep~(1D) spectra cross-correlated with the DRS G2 template adapted to our \gr~wavelength range (as done in Sect.~\ref{res:line_width_G2}); and on the other hand the data produced by the HARPS-North DRS based on the original G2 template and the corresponding 2D spectra of \dcep. The DRS cross-correlates each order of the 2D spectrum with the default G2 template before summing the CCFs over all the orders to obtain the average CCF, which is then fitted by a Gaussian. We display the results in Fig.~\ref{fig:comp_G2}. For the four compared observables (Gaussian \vr, Gaussian FWHM, BIS and CCF depth), we obtain a Pearson correlation coefficient between 0.99 and 1, showing the robustness of our \texttt{s1d}-based approach. A linear regression of the two Gaussian \vr~time series leads to a slope very close to 1 ($= 1.006 \pm 0.002$) and a non-significant zero-point. The differences in terms of CCF depth and FWHM amplitudes remain small (by $<$2 and $<$4\%, respectively) and probably originate in the fact that the DRS CCF computation is done on a wider wavelength interval than our \gr~range (encompassing all orders, \ie~from $\sim$3900 to $\sim$6900 \AA). The larger differences in terms of BIS amplitudes can be explained by the different BIS definitions between this study and the DRS. Overall, this confirms that using the \texttt{s1d} spectra with a S/N lower threshold (Sect.~\ref{method:spectra}) is a valid approach.

\section{New orbital parameters}\label{appdix:bin}

We detail here the computation of the companion orbital parameters to two of our Cepheid targets, SU~Cyg and V496~Aql. First, we combined for each target our \vr~data with previous \vr~time series from the literature, to obtain a time baseline roughly of the same order than the orbital period. Next, we used the {\it yorbit} tool \citep{segransan11} to model simultaneously the orbital and pulsation curves with Keplerian models. {\it yorbit} is based on a Levenberg-Marquardt algorithm that allows to fit \vr~data with Keplerian models (among others), after selecting the values with a genetic algorithm. We consider using a Keplerian fit to model the pulsation curve as a valid first approximation in the case of these two Cepheids as they are fundamental pulsators. The orbital parameters are detailed in Table~\ref{tab:bin_orbit}.

\renewcommand{\arraystretch}{1.25}
\begin{table}[t!]
\caption{New orbital parameters of two binary Cepheids.}
\label{tab:bin_orbit}
\begin{center}
\begin{tabular}{l c | c c}\\
\hline
\hline
Orbital     &   Unit      & SU~Cyg               & V496~Aql \\
parameters  &             &                      &          \\
\hline
$P_{\rm orb}$ & day         &  $538.5 \pm 4.2$     & $1351.5 \pm 5.8$   \\
T$_{\rm p}$   & MJD        &  $54744.6 \pm 17.4$  & $57537.5 \pm 12.1$    \\
$e$         &             &  $0.79 \pm 0.05$     & $0.45 \pm 0.04$    \\
$\omega$    & (\degr)     &  $168.7 \pm 8.2$     & $-68 \pm 4.1$   \\
$K$         & \kms        &  $30.59 \pm 0.82$    & $9.5 \pm 0.46$    \\
\hline
$M_{\rm 1}$~($\dagger$)   & \Msun    &   $4.7 ^{(1)}$        &  $5.6 ^{(1)}$           \\
$\pi_{\rm p}$~($\dagger$) & mas      & $1.27 \pm 0.86 ^{(2)}$&  $0.94 \pm 0.05 ^{(3)}$            \\
\hline
$M_{\rm 2}\sin{i}$ & \Msun         &  $2$                 &   $1.4$            \\
$a$              &   au          &   $2.45$             &   $4.57$             \\

\hline
\hline
\end{tabular}
\end{center}
$\dagger$ Parameters assumed from the literature.
{\it (1)} \cite{evans15}.
{\it (2)} HIPPARCOS parallax from \cite{vanleeuwen07}.
{\it (3)} {\it Gaia} DR2 parallax \citep{gaia18}.
\end{table}
\renewcommand{\arraystretch}{1}

 \paragraph{SU~Cyg --} We combined our SOPHIE~data with \vr~taken from \cite{gorynya98}. Our best {\it yorbit} model corresponds to an orbital period $P_{\rm orb}$~$\sim$539~days (or $\sim$1.5~year), in agreement with \cite{evans15}, and a $\sim$0.8~high eccentricity. Assuming the primary mass from \cite{evans15} and the HIPPARCOS parallax (Table~\ref{tab:bin_orbit}), we deduce a minimal mass of 2~\Msun~and a semi-major axis (sma) of $\sim$2.5~au for the secondary. In comparison, \cite{evans15} found a true secondary mass of 3.2~\Msun. \cite{kervella18} reported an orbital period of 549 days (\ie~close to our {\it yorbit} value) but with an eccentricity twice smaller and a companion true secondary mass of 4.7~\Msun. Our high eccentricity value may be an artefact induced by the Keplerian modelling of an uneven \vr~sample.

 \paragraph{V496~Aql --} We combined our HARPS, SOPHIE and CORALIE data with \vr~taken from \cite{gorynya98}, \cite{storm11+} and \cite{groen13}. Our best model corresponds to an orbital period of $1352 \pm 6$~days or $\sim$3.7~years, \ie~somewhat longer than the 2.9~year period of \cite{evans15}. \cite{groen13} reported a significantly smaller orbital period ($\sim$1066~days), but with a null-eccentricity model, which probably explains the difference. Assuming the primary mass of these authors and the {\it Gaia} DR2 parallax, we deduce a companion minimal mass of 1.4~\Msun~and a companion sma of~$\sim$4.6 au. In comparison, \cite{evans15} found a true secondary mass of 1.9~\Msun. 

\end{appendix}

\end{document}